\documentclass[notitlepage,twocolumn,aps,prl,longbibliography,superscriptaddress]{revtex4-2}

\usepackage{amsmath}
\usepackage{changes}
\usepackage{graphicx}
\usepackage[space]{grffile} 
\usepackage{gensymb}
\usepackage{bm}
\usepackage{float}
\usepackage{amssymb,amsfonts,amsmath,amsbsy,bm,t1enc,latexsym}

\usepackage{changes}
\usepackage{soul}
\usepackage[super]{nth}
\usepackage[normalem]{ulem}
\usepackage{siunitx} 
\usepackage[
bookmarks=true,
colorlinks=true,
linkcolor=blue,
urlcolor=blue,
citecolor=blue,
pdftex,
bookmarks=true,
linktocpage=true, 
hyperindex=true
]{hyperref}
\usepackage{hyperref}
\usepackage{siunitx}
\usepackage[super]{nth}

\newcommand{\SiN}[0]{$\mathrm{Si}_3\mathrm{N}_4$~}

\begin{document}	
	
\title{Photonic-electronic integrated circuit-based coherent LiDAR engine}
	
	\author{Anton Lukashchuk}
	\thanks{These authors contributed equally}	
	\affiliation{Laboratory of Photonics and Quantum Measurements (LPQM), Swiss Federal Institute of Technology Lausanne (EPFL), CH-1015 Lausanne, Switzerland}
	
	\author{Halil Kerim Yildirim}
	\thanks{These authors contributed equally}	
	\affiliation{Advanced Quantum Architecture Laboratory (AQUA), Switzerland, Swiss Federal Institute of Technology Lausanne (EPFL), CH-2002 Neuchâtel, Switzerland}
	
	\author{Andrea Bancora}
	\affiliation{Laboratory of Photonics and Quantum Measurements (LPQM), Swiss Federal Institute of Technology Lausanne (EPFL), CH-1015 Lausanne, Switzerland}
	
	\author{Grigory Lihachev}
	\affiliation{Laboratory of Photonics and Quantum Measurements (LPQM), Swiss Federal Institute of Technology Lausanne (EPFL), CH-1015 Lausanne, Switzerland}
	
	\author{Yang Liu}
	\affiliation{Laboratory of Photonics and Quantum Measurements (LPQM), Swiss Federal Institute of Technology Lausanne (EPFL), CH-1015 Lausanne, Switzerland}	
	
	\author{Zheru Qiu}
	\affiliation{Laboratory of Photonics and Quantum Measurements (LPQM), Swiss Federal Institute of Technology Lausanne (EPFL), CH-1015 Lausanne, Switzerland}	
	
	\author{Xinru Ji}
	\affiliation{Laboratory of Photonics and Quantum Measurements (LPQM), Swiss Federal Institute of Technology Lausanne (EPFL), CH-1015 Lausanne, Switzerland}	
	
	\author{Andrey Voloshin}
	\affiliation{Laboratory of Photonics and Quantum Measurements (LPQM), Swiss Federal Institute of Technology Lausanne (EPFL), CH-1015 Lausanne, Switzerland}	
	
	\author{Sunil A. Bhave}
	\affiliation{OxideMEMS Lab, Purdue University, 47907 West Lafayette, IN, USA}
	
	\author{Edoardo Charbon}
	\email{edoardo.charbon@epfl.ch}
	\affiliation{Advanced Quantum Architecture Laboratory (AQUA), Switzerland, Swiss Federal Institute of Technology Lausanne (EPFL), CH-2002 Neuchâtel, Switzerland}
	
	\author{Tobias J. Kippenberg}
	\email{tobias.kippenberg@epfl.ch}
	\affiliation{Laboratory of Photonics and Quantum Measurements (LPQM), Swiss Federal Institute of Technology Lausanne (EPFL), CH-1015 Lausanne, Switzerland}
	
	\date{\today}
	
	\pacs{}
	
	\maketitle



\textbf{  
Microelectronic integration is a key enabler for the ubiquitous deployment of devices in large volumes ranging from MEMS and imaging sensors to consumer electronics. Such integration has also been achieved in photonics, where compact optical transceivers for data centers employ co-integrated photonic and electronic components.
Chip-scale integration is of particular interest to coherent laser ranging i.e. frequency modulated continuous wave (FMCW LiDAR), a perception technology that benefits from instantaneous velocity and distance detection, eye-safe operation, long-range and immunity to interference.
While major progress in integrated silicon photonics based beam-forming has been made, full wafer-scale integration of this technology has been compounded by the stringent requirements on the lasers, requiring high optical coherence, low chirp nonlinearity and requiring optical amplifiers. 
Here, we overcome this challenge and demonstrate a photonic-electronic integrated circuit-based coherent LiDAR engine, that combined all functionalities using fully foundry-compatible wafer scale manufacturing. 
It is comprised of a micro-electronic based high voltage arbitrary waveform generator, a hybrid photonic circuit based tunable Vernier laser with piezoelectric actuators, and an erbium-doped waveguide optical amplifier - all realized in a wafer scale manufacturing compatible process that comprises III-V semiconductors, \SiN silicon nitride photonic integrated circuits as well as 130-nm SiGe BiCMOS technology.
We used the LiDAR engine to conduct 3D ranging experiments with chirps of 2~GHz excursion and 50~kHz sweep rate, operating at over 20~mW optical power, and performing scene mapping at a distance of 10 meters. The source is a turnkey, linearization-free, and can serve as a 'drop-in' solution in any FMCW LiDAR, that can be seamlessly integrated with an existing focal plane and optical phased array LiDAR approaches, constituting a missing step towards a fully chip-scale integrated LiDAR system.
}



\section{Introduction}

Laser ranging (LiDAR) is widespread perception technology that is rapidly developing using recent progress in silicon photonics \cite{Rogers2020, Zhang2022, Poulton2022}. LiDAR is ubiquitous in robotics, spatial mapping, and AR/VR applications and gained popularity in the early 2000s as a key enabler of autonomous vehicles in urban environments, a goal highlighted by DARPA Grand Challenges \cite{Urmson2008}. 
Widely employed in the early 2000's time-of-flight sensors, which measure the arrival time of reflected pulses, relied on available legacy 900~nm diode lasers and silicon detectors.
Another type of LiDAR is frequency modulated continous wave (FMCW) LiDAR \cite{Bostick1967,Pierrottet2008}, which maps distance and velocity of an object to frequency. 
This method, an optical analogue of coherent RADAR, utilizes optical self heterodyne detection of a frequency-chirped continuous-wave light reflected from a target with its replica, that serves as the local oscillator (LO). 
In contrast to the time-of-flight approach, coherent ranging allows for instantaneous velocity measurement via the Doppler frequency shift, quantum noise limited detection enabled by heterodyne detection with sufficient LO power, eye-safe operation at low average powers, immunity to ambient light sources, and low-bandwidth receiver electronics (100s of MHz) capable of providing cm-level resolution mainly dependent on frequency excursion of the transmitted chirp. However, the cost and bulky size of individual LiDAR components and their assembly still preclude the wide adoption of ranging sensors. 
 
\noindent Frequency-modulated continuous wave (FMCW) LiDAR, in particular, requires multiple building blocks including a frequency-agile laser, driving electronics, scanning optics, passive components (grating couplers, switching network), and detectors.
A variety of recent work attempted to integrate coherent LiDAR components on chip. 
Martin et al. demonstrated a silicon photonic circuit with integrated detectors, waveform calibration and switching network for passive beam scanning capable of 60~m coherent ranging at 5~mW output power \cite{Martin2018}.
Poulton et al. presented an optical phased array (OPA) based FMCW 3D ranging up to 10~m \cite{Poulton2019}. 
The same group demonstrated nearly centimeter scale OPA aperture with 8192 elements achieving 100$^o\times$17$^o$ field of view \cite{Poulton2022}.
Rogers et al. developed a focal plane array (FPA) 3D LiDAR on a silicon chip with photonic-electronic monolithic integration of 512 pixel coherent receiver array \cite{Rogers2020}.

\noindent The aforementioned approaches are CMOS compatible, can be integrated with other passive or active optical components and are scalable, i.e., support a further increase in the number of pixels and field of view. 
However, these prior demonstrations all used \emph{external} lasers coupled via fiber, off-the-shelf driving electronics, and bulk fiber based erbium-doped amplifiers for signal or reflection amplification and bulk modulators (Refs. \cite{Poulton2019, Rogers2020}) - significantly compounding full integration. 
A fully integrated FMCW LiDAR will require to address the remaining integration,
and replace these building blocks with their photonic integrated circuit based counterparts.
Tackling the issue of discrete external components, Isaac et al. fabricated an integrated transceiver module on the InP platform \cite{Isaac2019}, but no coherent ranging functionality was performed. 
Ref. \cite{Sayyah2022} showed a fully integrated coherent LiDAR on chip, though it is limited to single-pixel imaging only.
Here we overcome this challenge and report a fully photonic-electronic integrated circuits LiDAR engine.
It combines an integrated laser, high-voltage arbitrary waveform generator (HV-AWG) application-specific integrated circuit (ASIC), and chip-scale Erbium amplifier - which are all manufactured with foundry compatible wafer scale manufacturing.
First, we demonstrated a fully integrated HV-AWG in a standard 130-nm SiGe BiCMOS technology, which can be integrated with advanced electronics thanks to its small feature size while enabling tight assembly with photonic integrated circuits (PIC) technology for a miniaturized system. 
The ASIC uses a novel charge pump architecture, which generates arbitrary waveforms exceeding 20~V with no need for external high voltage supplies.
Further, we used a hybrid integrated Vernier laser based on a low-loss \SiN platform with fast piezoelectric actuators that supports >100~kHz sweep rate and >2~GHz frequency excursion.
Lastly, we employed an erbium-doped waveguide amplifier (EDWA) capable of providing 24~dB off-chip net gain \cite{Liu2022} to meet the optical power budget requirement and compensate for the losses caused by chip-to-chip coupling, switching network and local oscillator (LO) driving \cite{Martin2018, Rogers2020}. 
Using these photonic integrated circuit-based components we demonstrated ranging at 10~m with cm-level precision.
The combination of integrated laser, HV-AWG ASIC and chip-scale Erbium amplifier constitutes a plug-and-play coherent LiDAR source, which can be applied to existing silicon imaging 3D sensors \cite{Rogers2020,Zhang2022,Poulton2019} and pave a path towards a fully integrated coherent LiDAR system.

\begin{figure*}[!htbp]
	\includegraphics[width=\linewidth]{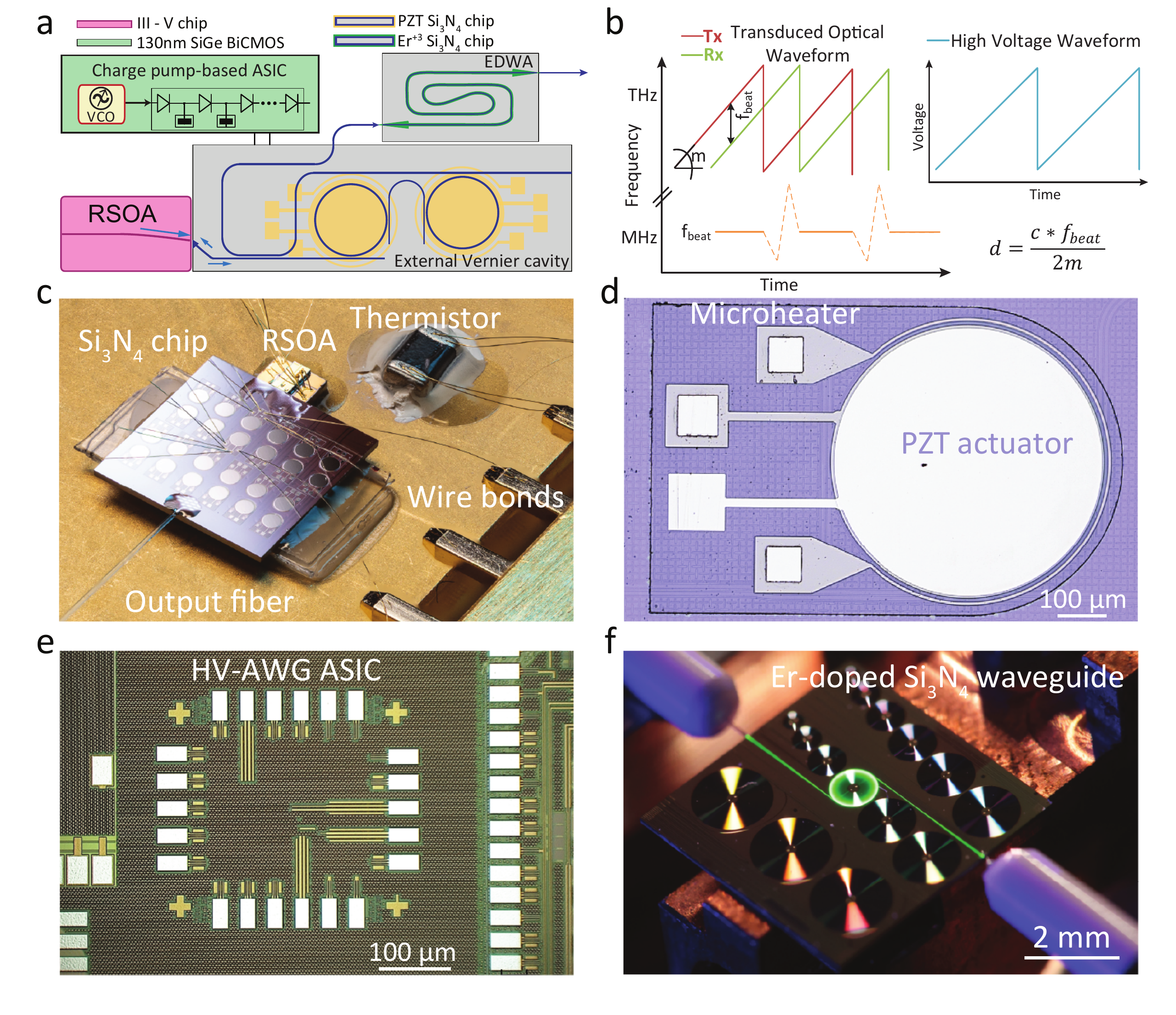} 
	\caption{\footnotesize{\textbf{Concept of photonic-electronic LiDAR source.}
	a)~Schematics of photonic-electronic LiDAR structure comprising a hybrid integrated laser source, charge-pump based HV-AWG ASIC, photonic integrated erbium-doped waveguide amplifier.
	b)~Coherent ranging principle. 
	c)~Packaged laser source. RSOA is edge coupled to \SiN Vernier filter configuration waveguide, whereas the output is glued to the fiber port. PZT and microheater actuators are wirebonded as well as butterfly package thermistor. 
	d)~Zoom in view of c) highlighting a microring with actuators.
	e)~Micrograph of the HV-AWG ASIC chip fabricated in a 130~nm SiGe BiCMOS technology. The total size of the chip is 1.17-1.07~mm$^2$.
	f)~The Erbium-doped waveguide is optically excited by a 1480~nm pump showing green luminescence due to the transition from a higher lying energy level to the ground state.  
	}	
	}
	\label{fig_concept}
\end{figure*}

\section*{Results}

\subsection{Photonic-electronic LiDAR source.}
The photonic-electronic LiDAR consists (cf. Fig.~\ref{fig_concept}a) of three main building blocks: laser, ASIC and on-chip amplifier. 
Generally, a distributed feedback laser (DFB) is used as the light source in FMCW LiDAR implementations \cite{Amann1992, Karlsson1999, Zhang2019}. 
While DFB lasers offer tunability with MHz actuation bandwidth and excursions up to 100s of GHz, they suffer from the need for continuous feedback for chirp linearity \cite{Behroozpour2016} -- the linearity-phase noise and wide tunability trade-off inherent to conventional lasers \cite{Tran2019}.
Recent advances in integrated photonics have enabled to achieve fiber laser coherence, by using self-injection locking of a DFB laser to ultra-low propagation loss integrated photonic circuits based on silicon nitride \cite{Jin2021}.
By endowing such circuits with piezoelectrical MEMS actuators, it has been possible to achieve both high coherence as well as fast tuning with low nonlinearity \cite{Lihachev2022}, allowing linearization free FCMW LiDAR -- yet require large voltage driving.
Other approaches based on electro-optic integrated photonic laser feedback circuits using $\chi^{(2)}$ materials, such as LiNbO$_3$ or BTO \cite{Li2022, Snigirev2022, Li2022a} can lead to even faster frequency chirps at lower voltage, but presently exhibit far lower laser coherence.

\noindent In our work, we employ an external cavity hybrid integrated laser (cf. Fig. \ref{fig_concept}c) operating at 1566~nm that includes a reflective semiconductor optical amplifier (RSOA) edge-coupled to a \SiN photonic integrated circuit with a microresonator-based Vernier filter  \cite{Rees2020}. 
This approach has the distinct advantage of using cost-effective III-V based RSOA, that does not require gratings as in the case of DFB.
Two microresonators with loaded cavity linewidths of 200~MHz slightly differ in free spectral range (FSR), i.e., 96.7 and 97.9~GHz with Vernier frequency of 8.7 THz.
Both rings have integrated microheaters realized in the bottom Pt electrode layer.  
We aligned the pair of resonances using a microheater and obtained up to 20$\%$ reflection back to the RSOA, using the feedback circuit depicted in Fig. \ref{fig_concept}a.
Piezoelectric PZT actuators are heterogeneously integrated on top of the photonic integrated microrings to perform fast actuation via stress-optic effect for rapid laser frequency tuning.
Platinum electrodes (cf. Fig. \ref{fig_concept}d) match the resonator radius to maximize the stress-optic tuning efficiency, attaining $\sim$130~MHz/V of frequency tuning.
When coupling the hybrid MEMS-based circuit to an RSOA, we observe lasing with an output power of 3~mW, side mode suppression of 50~dB, and featuring frequency noise of 10$^4$~Hz$^2$/Hz at 10~kHz offset frequency, reaching the white noise floor of 127 Hz$^2$/Hz at 6~MHz offset \cite{Bancora2022}. 
For better stability the entire assembly is packaged within a butterfly 14-pin package and placed on a Peltier element, and light is coupled to a SMF output fiber.
The RSOA, Peltier element, thermistor, all microheaters, and PZT actuators are all connected to the butterfly pins using electrical wirebonding.
The hybrid packaging allows for turnkey laser operation, reduces laser frequency noise at offsets below 1 kHz and maintains the tuning performance after waveform predistortion and linearization.

\noindent Operating the laser in FMCW mode (cf. Fig \ref{fig_concept}b) requires frequency tuning over several GHz, which for PZT integrated actuators necessitates voltages above 10~V, not achievable with conventional CMOS electronics.
To overcome this we designed and fabricated, a HV-AWG integrated circuit (see Fig. \ref{fig_concept}e) that generates a 15~V sawtooth waveform that drives the PZT actuators while being supplied with only 3.3~V. 
The electrical waveform is then transduced to the optical domain, resulting in a >2~GHz optical chirp excursion. 
FMCW can be implemented in various ways. 
Typically, it uses a triangularly chirped waveform \cite{Feneyrou2017}, but can also use random phase code modulation \cite{Axelsson2004}.
Fig. \ref{fig_concept}b shows the chirp waveform employed in our experiments. 
Linearly frequency-chirped laser light is split in local oscillator path and signal path, with the signal path sent to the target. 
The reflected light is then mixed with the local oscillator and measured on a balanced photodiode. 
The detected range is proportional to the beatnote frequency inferred from the short-time Fourier transform of the recorded heterodyne signal. 
It is inversely proportional to the waveform chirp rate $m$ - the ratio of the excursion and sweeping period. 

\noindent Finally, we employed a chip-scale integrated EDWA (cf. Fig. \ref{fig_concept}f) to amplify the laser to >20~mW optical power to meet the power requirements for robust and long-range coherent ranging \cite{Wang1982}.
Typically integrated LiDAR systems require $>$100~mW of optical power to compensate for coupling losses and be able to emit 10s~mW at the aperture \cite{Martin2018, Poulton2022}.  
The EDWA was implemented using an on-chip 21-cm-long \SiN spiral waveguide doped with high-concentration Erbium ions ($3.25\times10^{20} \mathrm{ions/cm}$) through a high-energy (up to 2 MeV) ion implantation process \cite{polman_optical_1991, Liu2022}.
The doped Erbium ions can be optically pumped to the excited state and allow for amplification stemming from the stimulated transition to the ground state.
The EDWA provides linear and low noise optical amplification to the frequency modulated optical waveform due to the slow gain dynamics (millisecond lifetime) and the small emission cross section of Erbium ions \cite{pollnau_rare-earth-ion-doped_2015}.

\subsection{High-voltage Arbitrary Waveform Generator ASIC.}
\begin{figure*}[!htbp]
	\includegraphics[width=\linewidth]{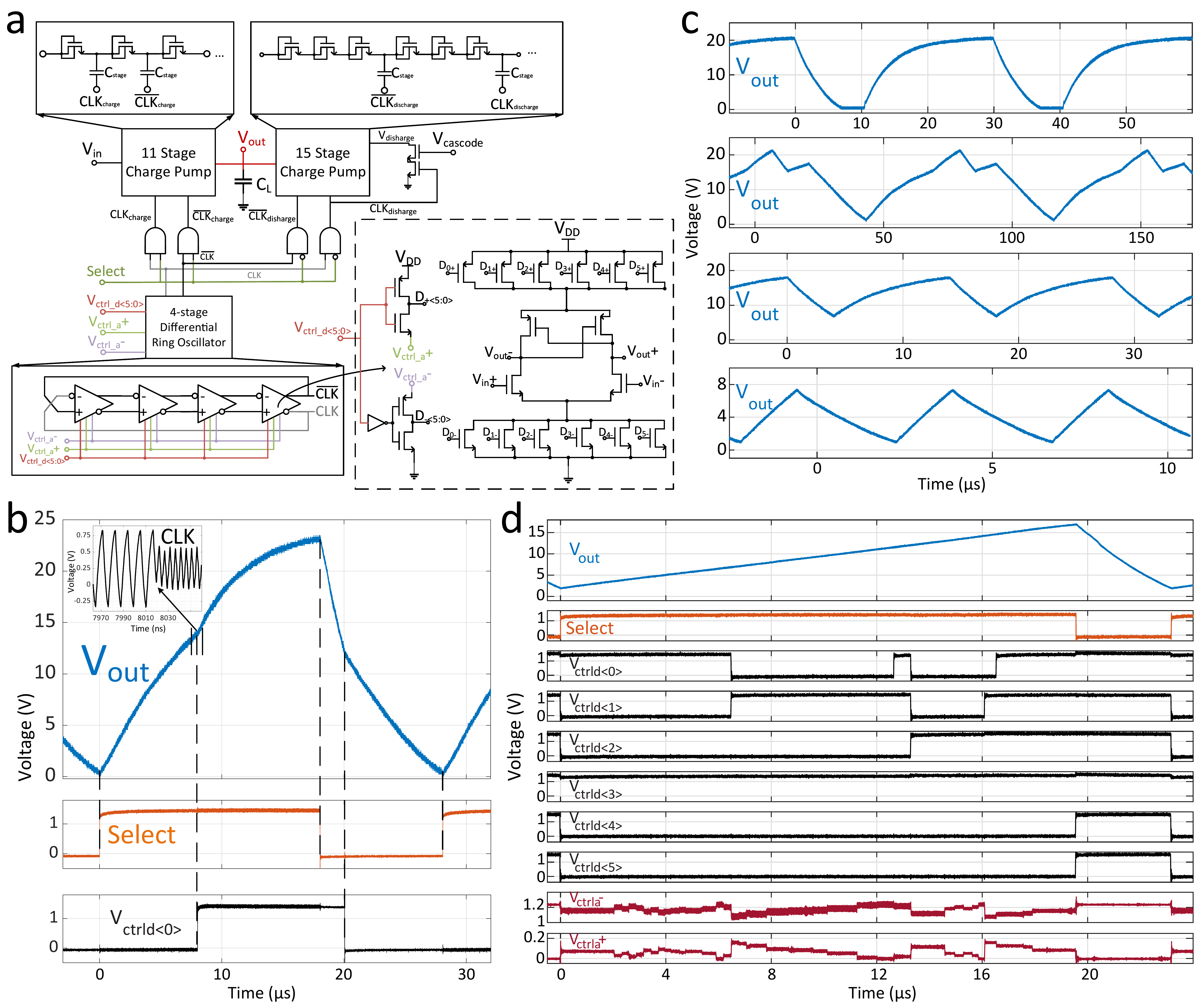}
	\caption{\footnotesize{\textbf{High-voltage arbitrary waveform generator integrated circuit, fabricated in a 130-nm SiGe BiCMOS technology. }
	a)~Schematics of the integrated circuit consisting of a 4-stage voltage-controlled differential ring oscillator which drives charge pump stages to generate high-voltage arbitrary waveforms.
	b)~Principles of waveform generation demonstrated by the output response to the applied control signals in the time domain. Inset shows the change in oscillation frequency in response to a frequency control input, from 88 MHz to 208 MHz, which modifies the output waveform.
	c)~Measured arbitrary waveforms generated by the ASIC with different shapes, amplitudes, periods and offset values.
	d)~Generation of the linearized sawtooth electrical waveform used in LiDAR measurements. Digital and analog control signals are modulated in the time domain to fine-tune the output.
	}
	}
	\label{fig_charge_pump}
\end{figure*}

Arbitrary waveform generation at high voltages is desired to drive various devices including ultrasound transducers \cite{Lee2021,Jung2018}, piezoactuators \cite{Lihachev2022}, neurostimulators \cite{Sooksood2011}, single-photon avalanche diodes (SPADs) \cite{Zhang2015}. 
HV-AWGs are usually provided as a single or even multiple discrete components \cite{Dragonas2015,Lihachev2022}, which are generally challenging to be integrated due to their incompatibility with technologies supporting advanced electronics.
We demonstrate a novel architecture which can generate high-voltage arbitrary waveforms using a standard CMOS technology supplied at 3.3~V. 
Fig. \ref{fig_charge_pump}a shows the schematic block diagram of the IC. 
The ASIC consists of a voltage-controlled ring oscillator (VCRO), which drives the clocks of a series of Dickson charge pump stages. 
The oscillation frequency can be controlled externally to modify the waveform.  
The 11-stage charge pump generates the output waveform rising edges, whereas the 15-stage charge pump generates falling edges.
The two charge pump blocks operate in a complementary fashion, with the ‘Select’ signal controlling whether the output voltage increases or decreases. 
The clocks are applied only to the charging/discharging block during one half cycle when the output voltage rises/falls. 
The oscillator is designed to have a wide frequency range and a high frequency resolution, so as to achieve fine-tuning capability while controlling the waveform. 
The unit cell of the four-stage ring oscillator has six geometrically sized pairs of NMOS and PMOS (n-type/p-type metal-oxide-semiconductor) transistor loads to control the unit delay. 
The inputs, $V_\mathrm{\mathrm{ctrld}<5:0>}$, can be digitally switched to turn on a pair of NMOS and PMOS loads each, where $V_\mathrm{ctrld<5>}$ corresponds to the load with the highest width per length. 
Using digital inputs, the gate voltage of the PMOS loads is connected to $V_\mathrm{ctrla+}$ and the NMOS loads to $V_\mathrm{ctrla-}$, respectively. 
These two gate voltages are controlled differentially in an analog fashion, where their sum equals $V_\mathrm{DD}$ to fine tune the oscillation frequency. 
Operating at a supply of 1.2 V, the frequency of the designed oscillator can be set in the range of 6 MHz to 350 MHz, with a tuning control of 2$\%$ at lower frequencies and 0.5$\%$ at higher frequencies employing 10 mV steps for $V_\mathrm{ctrla+}$ and $V_\mathrm{ctrla-}$.

\noindent Fig. \ref{fig_charge_pump}b illustrates the principles of waveform generation.
The output voltage of a charge pump in the time domain shows an exponential response on a capacitive load, where the rise-time depends on the frequency of the applied clocks \cite{Palumbo2010}. 
Therefore, one can tune the time-domain waveform by changing the clock frequency at pre-determined time points. 
We can set a digital control input of the VCRO high to decrease the waveform rise time by increasing the clock oscillation frequency, from approximately 88 MHz to 208 MHz. 
To start the falling edge, we switch 'Select' signal low, and the fall time is controlled in the same manner using the oscillator inputs. 
This allows the HV-AWG ASIC to generate output waveforms with a peak voltage of more than 20 V. 
The output waveform has a period similar to ‘Select’; ‘Select’ duty cycle also sets the highest and lowest output voltage values. 
We can produce waveforms with different shapes, amplitudes, frequencies and offset values when operating the circuit with different VCRO input sequences, shown in \ref{fig_charge_pump}c. 
Fig. \ref{fig_charge_pump}d shows the generated 45~kHz sawtooth waveform used in our FMCW LiDAR experiments. 
We linearized the exponential response of the charge pump to obtain a sawtooth waveform. 
The voltage control inputs are modified in time to gradually decrease the rise-time, by changing the VCRO frequency within the range of 30-80 MHz. 
The digital voltages, $V_\mathrm{ctrld<5:0>}$, allow a coarse control of the VCRO causing too abrupt changes in the waveform. 
The analog voltages, $V_\mathrm{ctrla+}$ and $V_\mathrm{ctrla-}$, are used in conjunction at smaller time steps which allows fine-tuning of the optical waveform for higher linearity.  

\subsection{Electro-optic transduction and linearity.}

\begin{figure*}[!htbp]
	\includegraphics[width=0.98\linewidth]{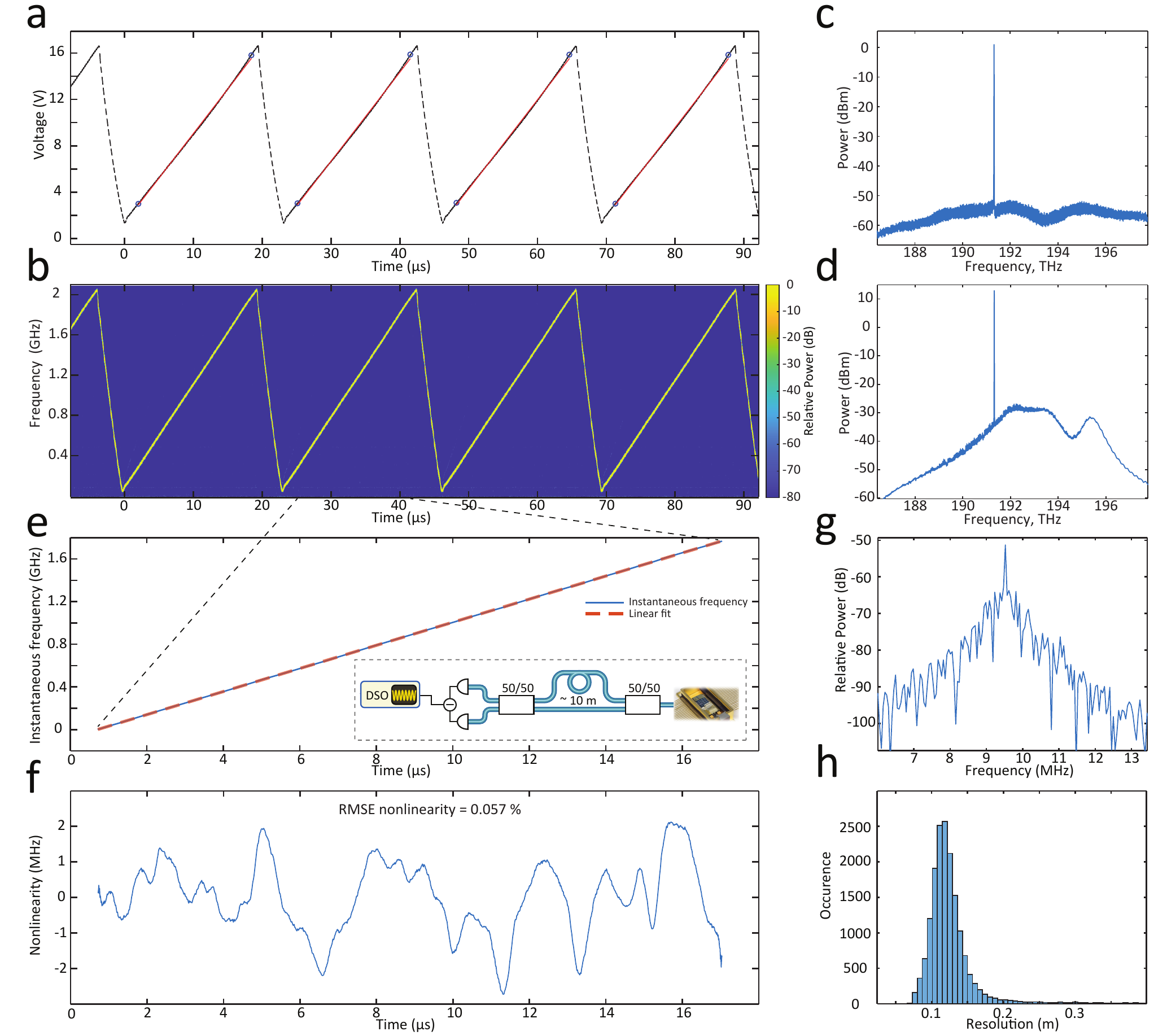}
	\caption{\footnotesize{\textbf{Photonic integrated LiDAR engine electro-optical transduction and linearity.} 
	a)~Electrical waveform generated by the ASIC. Blue circles highlight the segment of $\sim$16~$\mu$s used for ranging and linearity analysis. The red curve is a linear fit to the given segment. 
	b)~Time-frequency map of the laser chirp obtained via heterodyne detection with auxiliary laser. RBW is set to 10~MHz.  
	c)~Optical spectrum of Vernier laser output featuring 50~dB side mode suppression ratio.
	d)~Optical spectrum after EDWA with >20~mW optical power.
	e)~Instantaneous frequency of the optical chirp obtained via delayed homodyne measurement (left bottom inset: experimental setup). The red dashed line corresponds to the linear fit. The excursion of the chirp equates to 1.78~GHz over 16~$\mu$s period.
	f)~Nonlinearity of the laser chirp inferred from e). RMSE nonlinearity equates to 0.057$\%$ with the major chirp deviation from the linear fit lying in the window $\pm$2~MHz.  
	g)~The frequency beatnote in the delayed homodyne measurement corresponds to the reference MZI delay $\sim$10~m. The 90$\%$ fraction of the beatnote signal is taken for the Fourier transformation.   
	h)~LiDAR resolution inferred from the FWHM of the MZI beatnotes over >20000 realizations. Most probable resolution value is 11.5~cm while the native resolution is 9.3~cm corresponding to 1.61 GHz (90$\%$ of 1.78~GHz) 
	}
	}
	\label{fig_eo_trans}
\end{figure*}

Fig. \ref{fig_eo_trans}a demonstrates the electrical waveform generated by the ASIC. 
The same sawtooth signal was applied to both piezoactuators of the Vernier laser.
The heterodyne measurement carried out with an auxiliary laser shows the time-frequency map of the laser chirp (cf. Fig. \ref{fig_eo_trans}b).  
The 15~V electrical signal resulted in >2~GHz optical frequency excursion over 23 $\mu$s period. 
The chirp $m$ parameter of the up-swing used for the ranging equates to $\sim$110~THz/s. 
It ultimately determines the detected beatnote frequency $f_\mathrm{beat}$ to distance $d$ mapping via $d = c/2m \times f_\mathrm{beat}$ where $c$ is the speed of light. 
In our experiment 1~m of range maps to $\sim$1~MHz frequency beatnote for the laser sweep parameters described above.  

\noindent For the long range and robust measurement, FMCW LiDAR requires high chirp linearity of the optical waveform \cite{Zhang2019}. 
We iteratively linearized the optical waveform employing a delayed homodyne detection method \cite{Ahn2007}. 
We calculated the instantaneous frequency of the chirp (cf. Fig. \ref{fig_eo_trans}e) via Hilbert transformation of the beatnote electric signal. 
The ASIC architecture allows for fine tuning the waveform in an arbitrary fashion, therefore we optimized the frequency control inputs of the ASIC to minimize the root mean square error RMSE of the instantaneous frequency. 
We used a 16~$\mu$s up-rise segment of the chirp for linearity analysis.

\noindent Fig \ref{fig_eo_trans}f depicts the chirp nonlinearity or instantaneous frequency deviation from the fitted linear line. 
The major part of the deviation lies within $\pm$2~MHz window and exhibits a total nonlinearity of $<$0.1$\%$. 
While we optimized the optical waveform, the voltage ramp appeared to have 0.35$\%$ linearity and 0.05~V RMSE at an overall 15~V voltage excursion due to the non-ideal electro-optic transduction. 
We assume the electrical waveform noise limits the linearity of the optical chirp. 
The VCRO has $\sim$1000 oscillations per waveform period, with one step output voltage increase occurring at each cycle with charge pumping. 
The steps change in the range of 20 mV at low output voltages down to 8~mV at higher output voltages. 
This imposes a limit on the linearity of 0.05$\%$ for 15~V chirps due to quantization error.

\noindent The Fourier transform of the delayed homodyne detection is shown in Fig. \ref{fig_eo_trans}g. 
The full width at half maximum (FWHM) of the beatnote determines the resolution of the LiDAR. 
The beatnote linewidth is nearly Fourier transform limited featuring 60~kHz width.
Fig. \ref{fig_eo_trans}h presents statistics over $2\times10^4$ measurements. 
It indicates 11.5~cm resolution $\Delta R$ (most probable value) while the native resolution for $B=$1.6~GHz (90$\%$ fraction of 1.78~GHz) excursion chirp equates to 9.3~cm following $\Delta R = c/2B$.

\begin{figure*}[!htbp] 
	\includegraphics[width=\linewidth]{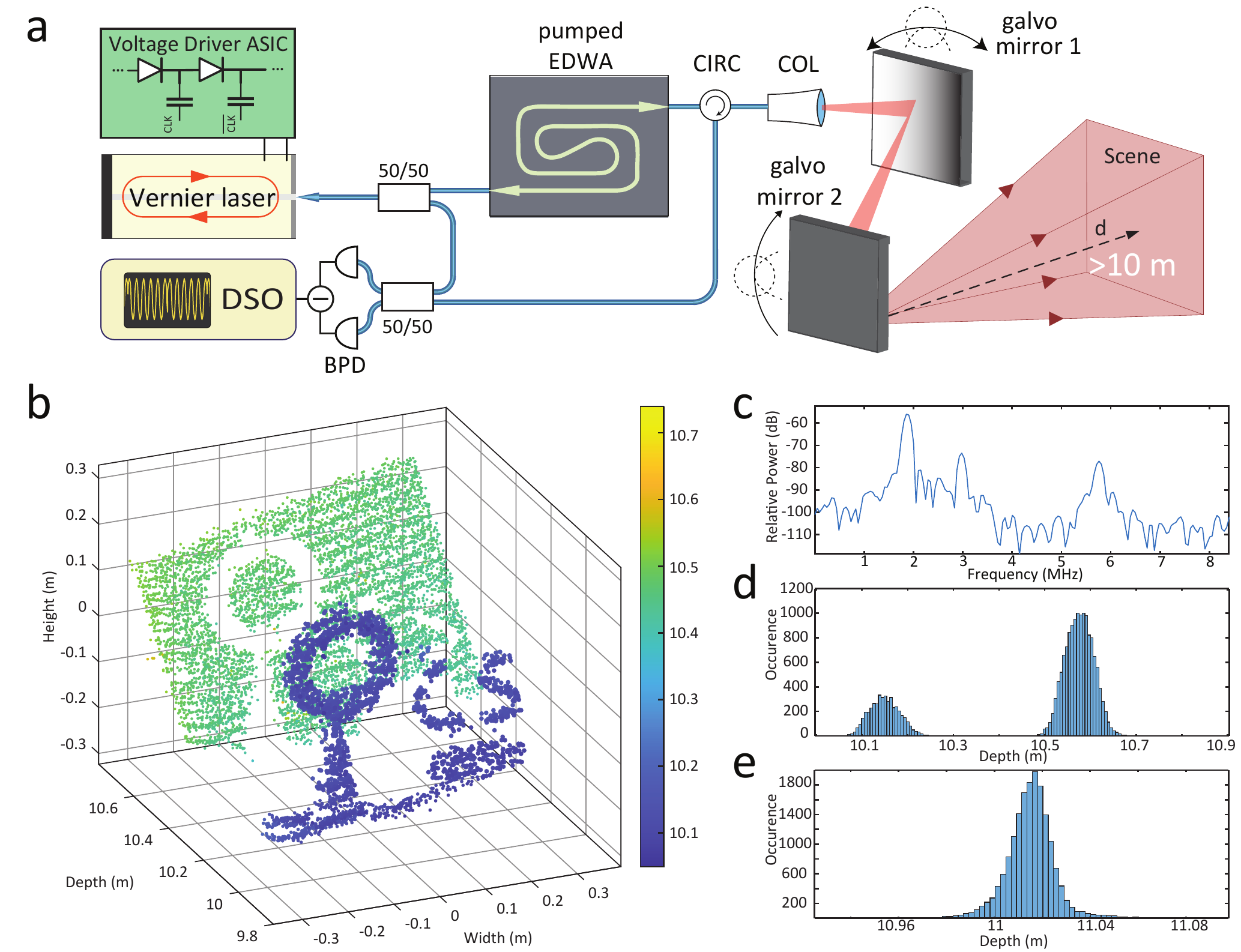}
	\caption{\footnotesize{\textbf{Ranging experiment.}
	a)~Schematics of the experimental setup for ranging experiments. The amplified laser chirp scans the target scene via a set of galvo mirrors. A digital sampling oscilloscope (DSO) records the balanced detected beating of the reflected and reference optical signals. CIRC - circulator, COL - collimator, BPD - balanced photodetector.
	b)~Point cloud consisting of $\sim10^4$ pixels featuring the doughnut on a cone and 'c', 's' letters as a target 10~m away from the collimator.
	c)~The Fourier transform over one period, highlighting collimator, circulator and target reflection beatnotes. Blackman–Harris window function was applied to the time trace prior to the Fourier transformation.
	d)~Detection histogram of b). 
	e)~Single point imaging depth histogram indicating 1.5~cm precision of the current setup.
	}
	}
	\label{fig_ranging}
\end{figure*}

\subsection{Optical ranging.}
Fig. \ref{fig_ranging}a illustrates the FMCW photonic-electronic LiDAR experimental setup. The Vernier laser was turnkey initiated, and the ASIC pre-configured waveform was subsequently applied to the laser piezo-actuators. 
The output light was first split into the signal and local oscillator paths. 
The EDWA chip provided 13~dB gain and amplified the signal up to 22~mW. 
The optical spectra before and after the amplification stage are depicted in Fig. \ref{fig_ranging}c,d, respectively. 
Further amplification is possible with double side pumping of the EDWA chip or by matching the Vernier lasing frequency to the maximum of the gain profile; Liu et al. demonstrated 24~dB off-chip amplification \cite{Liu2022}.
We employed a mono-static imaging setup with the same collimator operating as a transmitter and a receiver. 
Two galvo mirrors scanned the laser beam at 2~Hz vertical and 63~Hz horizontal rates.
An optical circulator separated the back-reflected light from the transmitted one, whereafter it was self-heterodyne mixed on a balanced photodetector. 
The digital oscilloscope sampled the photocurrent at 100 MS/s during 0.25~s acquisition time.  

\noindent Fig. \ref{fig_ranging}b shows the resulting point cloud consisting of $\sim10,000$ pixels. 
The target comprised a styrofoam doughnut and cone, 'c' and 's' paper letters, and a flat background placed 10~m away from the collimator (cf. photo of the target in SI). 
Every pixel was obtained by analyzing a single sawtooth period of 23 $\mu$s. 
First, the signal was time-gated to the up-swing fraction of the chirp, and then Fourier transformed. 
Fig. \ref{fig_ranging}c depicts a periodogram of the detected signal with the Blackman-Harris window applied. The chirp excursion was obtained from the self-homodyne measurement with an auxiliary interferometer.
We performed the Gaussian fitting of the target beatnote to infer the range estimate. 
We estimated the precision of our setup to be $\sim$1.5~cm derived from the statistics of a single point measurement over $\sim2\times10^4$ realizations.


\section*{Discussion}
In summary, we have demonstrated an integrated circuit-based coherent LiDAR engine. The system-level architecture represents a 'plug-and-play', 'drop-in' high-power frequency-agile laser with ASIC defined FMCW laser tuning.
We showed that a combination of a hybrid integrated Vernier ring laser with fast PZT actuators, a HV-AWG ASIC and an EDWA \SiN chip, attains >2~GHz frequency sweeps at 50~kHz rate with an output power of more than 20~mW.
Our LiDAR engine achieves 12~cm lateral ranging resolution with chirp nonlinearity of less than 0.1$\%$. 
Employing conventional 2D mechanical galvo mirror scanning, we demonstrated ranging at 10~m distance with 1.5~cm precision. 

\noindent The demonstrated ASIC architecture allows for high-voltage arbitrary waveform generation without using a high-voltage supply or RF power amplifier. 
The charge pump based design can generate output waveforms beyond the voltage rating of the transistors in the technology, therefore the driver can be implemented in advanced standard CMOS nodes. 
This is advantageous in terms of power consumption and computational processing capabilities of the system, but the manufacturing cost is typically higher. However, since the design has a small footprint with the possibility of having processing and high-voltage driver on a single CMOS chip, it supports further integration with the PIC assembly, allowing further integration of the LiDAR engine. 
SI Fig.1 shows the comparison of the proposed HV-AWG with other techniques in the literature, where integration, compactness and versatility are some of the advantages of the architecture. 
The chirp rate can be further increased from 50 kHz by reducing the parasitic capacitance at the ASIC output by minimizing the packaging or co-integration with the PIC.

\noindent Optical power amplification is commonly used in LiDAR demonstrations to achieve the required power levels, since the power output of a standalone laser is often insufficient.  
On-chip low-noise, high-power optical amplification has been recently made possible in Erbium-implanted \SiN photonic integrated circuits, showing competitive performance to widely deployed electrically pumped SOAs.
In contrast to SOAs with picosecond scale carrier lifetime, the Erbium-based amplifiers enable optical amplification of modulated optical signals with negligible gain nonlinearity or channel cross talk, and lower spontaneous emission noise, benefiting from the millisecond scale long excited state lifetime and much smaller emission cross-section.
Equally important, the EDWAs based on dielectric materials exhibit higher temperature stability than SOAs \cite{pollnau_rare-earth-ion-doped_2015}, which is critical for amplifying FMCW LiDAR signals. 
Temperature-induced index change, or phase shift, can cause undesirable instantaneous frequency drift and thus corrupt frequency-time linearity.
Although we have demonstrated separate integrated components, we note that the Vernier laser and the EDWA are based on the same \SiN material. 
The laser source and the amplifier can be simultaneously integrated on the same photonic chip due to possibility of selective area Er implantation, providing improved signal-to-noise ratio and higher output and low fabrication complexity.
We used a 1480~nm pump source for Er-ion excitation off-chip.
However, it could be hybrid integrated via edge-coupling \cite{Zhu2019} or photonic-wire bonded \cite{Blaicher2020}. 

\noindent Finally, we have presented a photonic-electronic integrated coherent ranging source. 
Comprising integrated laser, HV-AWG ASIC and on-chip amplifier, it could be readily applied to existing LiDAR approaches \cite{Rogers2020,Zhang2022,Poulton2019} replacing bulk components.

\newpage
\section*{Acknowledgements}
This material is based upon work supported by the Air Force Office of Scientific Research under Award No. FA9550-19-1-0250.  The work was further supported by the Swiss National Science Foundation (SNF) under contract No. 192293.  A.L. acknowledges support from the European Space Technology Centre with ESA Contract No. 4000133568/20/NL/MH/hm. This work also received funding from the EU H2020 research and innovation programme under the Marie Sklodowska-Curie grant agreement No. 101033663 (RaMSoM).

\section*{Data Availability Statement}
Relevant data, figures and analysis code will be published on \texttt{Zenodo} upon publication of the work.

\section*{Disclosures}
T.J.K. is a co-founder and shareholder of LiGenTec
SA, a foundry commercializing \SiN photonic integrated circuits.
T.J.K., A.V. are co-founders and shareholders of DEEPLIGHT SA and
A.B. is a shareholder of DEEPLIGHT SA, a start-up company
commercializing \SiN photonic integrated circuits based frequency
agile low noise lasers.

\section*{Author contributions}
H.K.Y. designed the ASIC. 
A.B. and A.V. performed hybrid packaging of the laser.
Y.L., Z.Q. and X.J. developed the EDWA chip.
A.L. and H.K.Y. performed the measurements with assistance from A.B. G.L. and Y.L.
A.L. performed the data analysis.
A.L. and H.K.Y. wrote the manuscript with input from all authors. 
T.J.K., E.C. and S.B. supervised the work. 


\begin{thebibliography}{39}%
\makeatletter
\providecommand \@ifxundefined [1]{%
 \@ifx{#1\undefined}
}%
\providecommand \@ifnum [1]{%
 \ifnum #1\expandafter \@firstoftwo
 \else \expandafter \@secondoftwo
 \fi
}%
\providecommand \@ifx [1]{%
 \ifx #1\expandafter \@firstoftwo
 \else \expandafter \@secondoftwo
 \fi
}%
\providecommand \natexlab [1]{#1}%
\providecommand \enquote  [1]{``#1''}%
\providecommand \bibnamefont  [1]{#1}%
\providecommand \bibfnamefont [1]{#1}%
\providecommand \citenamefont [1]{#1}%
\providecommand \href@noop [0]{\@secondoftwo}%
\providecommand \href [0]{\begingroup \@sanitize@url \@href}%
\providecommand \@href[1]{\@@startlink{#1}\@@href}%
\providecommand \@@href[1]{\endgroup#1\@@endlink}%
\providecommand \@sanitize@url [0]{\catcode `\\12\catcode `\$12\catcode
  `\&12\catcode `\#12\catcode `\^12\catcode `\_12\catcode `\%12\relax}%
\providecommand \@@startlink[1]{}%
\providecommand \@@endlink[0]{}%
\providecommand \url  [0]{\begingroup\@sanitize@url \@url }%
\providecommand \@url [1]{\endgroup\@href {#1}{\urlprefix }}%
\providecommand \urlprefix  [0]{URL }%
\providecommand \Eprint [0]{\href }%
\providecommand \doibase [0]{https://doi.org/}%
\providecommand \selectlanguage [0]{\@gobble}%
\providecommand \bibinfo  [0]{\@secondoftwo}%
\providecommand \bibfield  [0]{\@secondoftwo}%
\providecommand \translation [1]{[#1]}%
\providecommand \BibitemOpen [0]{}%
\providecommand \bibitemStop [0]{}%
\providecommand \bibitemNoStop [0]{.\EOS\space}%
\providecommand \EOS [0]{\spacefactor3000\relax}%
\providecommand \BibitemShut  [1]{\csname bibitem#1\endcsname}%
\let\auto@bib@innerbib\@empty
\bibitem [{\citenamefont {Rogers}\ \emph {et~al.}(2021)\citenamefont {Rogers},
  \citenamefont {Piggott}, \citenamefont {Thomson}, \citenamefont {Wiser},
  \citenamefont {Opris}, \citenamefont {Fortune}, \citenamefont {Compston},
  \citenamefont {Gondarenko}, \citenamefont {Meng}, \citenamefont {Chen} \emph
  {et~al.}}]{Rogers2020}%
  \BibitemOpen
  \bibfield  {author} {\bibinfo {author} {\bibfnamefont {C.}~\bibnamefont
  {Rogers}}, \bibinfo {author} {\bibfnamefont {A.~Y.}\ \bibnamefont {Piggott}},
  \bibinfo {author} {\bibfnamefont {D.~J.}\ \bibnamefont {Thomson}}, \bibinfo
  {author} {\bibfnamefont {R.~F.}\ \bibnamefont {Wiser}}, \bibinfo {author}
  {\bibfnamefont {I.~E.}\ \bibnamefont {Opris}}, \bibinfo {author}
  {\bibfnamefont {S.~A.}\ \bibnamefont {Fortune}}, \bibinfo {author}
  {\bibfnamefont {A.~J.}\ \bibnamefont {Compston}}, \bibinfo {author}
  {\bibfnamefont {A.}~\bibnamefont {Gondarenko}}, \bibinfo {author}
  {\bibfnamefont {F.}~\bibnamefont {Meng}}, \bibinfo {author} {\bibfnamefont
  {X.}~\bibnamefont {Chen}}, \emph {et~al.},\ }\bibfield  {title} {\bibinfo
  {title} {A universal 3d imaging sensor on a silicon photonics platform},\
  }\href {https://www.nature.com/articles/s41586-021-03259-y} {\bibfield
  {journal} {\bibinfo  {journal} {Nature}\ }\textbf {\bibinfo {volume} {590}},\
  \bibinfo {pages} {256} (\bibinfo {year} {2021})}\BibitemShut {NoStop}%
\bibitem [{\citenamefont {Zhang}\ \emph {et~al.}(2022)\citenamefont {Zhang},
  \citenamefont {Ding}, \citenamefont {Wang}, \citenamefont {Jiang},
  \citenamefont {Lou}, \citenamefont {Lu},\ and\ \citenamefont
  {Guo}}]{Zhang2022}%
  \BibitemOpen
  \bibfield  {author} {\bibinfo {author} {\bibfnamefont {G.}~\bibnamefont
  {Zhang}}, \bibinfo {author} {\bibfnamefont {Z.}~\bibnamefont {Ding}},
  \bibinfo {author} {\bibfnamefont {K.}~\bibnamefont {Wang}}, \bibinfo {author}
  {\bibfnamefont {C.}~\bibnamefont {Jiang}}, \bibinfo {author} {\bibfnamefont
  {J.}~\bibnamefont {Lou}}, \bibinfo {author} {\bibfnamefont {Q.}~\bibnamefont
  {Lu}},\ and\ \bibinfo {author} {\bibfnamefont {W.}~\bibnamefont {Guo}},\
  }\bibfield  {title} {\bibinfo {title} {Demonstration of high output power
  {DBR} laser integrated with {SOA} for the {FMCW} {LiDAR} system},\ }\href
  {https://doi.org/10.1364/OE.448993} {\bibfield  {journal} {\bibinfo
  {journal} {Optics Express}\ }\textbf {\bibinfo {volume} {30}},\ \bibinfo
  {pages} {2599} (\bibinfo {year} {2022})}\BibitemShut {NoStop}%
\bibitem [{\citenamefont {Poulton}\ \emph {et~al.}(2022)\citenamefont
  {Poulton}, \citenamefont {Byrd}, \citenamefont {Russo}, \citenamefont {Moss},
  \citenamefont {Shatrovoy}, \citenamefont {Khandaker},\ and\ \citenamefont
  {Watts}}]{Poulton2022}%
  \BibitemOpen
  \bibfield  {author} {\bibinfo {author} {\bibfnamefont {C.~V.}\ \bibnamefont
  {Poulton}}, \bibinfo {author} {\bibfnamefont {M.~J.}\ \bibnamefont {Byrd}},
  \bibinfo {author} {\bibfnamefont {P.}~\bibnamefont {Russo}}, \bibinfo
  {author} {\bibfnamefont {B.}~\bibnamefont {Moss}}, \bibinfo {author}
  {\bibfnamefont {O.}~\bibnamefont {Shatrovoy}}, \bibinfo {author}
  {\bibfnamefont {M.}~\bibnamefont {Khandaker}},\ and\ \bibinfo {author}
  {\bibfnamefont {M.~R.}\ \bibnamefont {Watts}},\ }\bibfield  {title} {\bibinfo
  {title} {Coherent {LiDAR} {With} an 8,192-{Element} {Optical} {Phased}
  {Array} and {Driving} {Laser}},\ }\href
  {https://doi.org/10.1109/JSTQE.2022.3187707} {\bibfield  {journal} {\bibinfo
  {journal} {IEEE Journal of Selected Topics in Quantum Electronics}\ }\textbf
  {\bibinfo {volume} {28}},\ \bibinfo {pages} {1} (\bibinfo {year}
  {2022})}\BibitemShut {NoStop}%
\bibitem [{\citenamefont {Urmson}\ \emph {et~al.}(2008)\citenamefont {Urmson},
  \citenamefont {Anhalt}, \citenamefont {Bagnell}, \citenamefont {Baker},
  \citenamefont {Bittner}, \citenamefont {Clark}, \citenamefont {Dolan},
  \citenamefont {Duggins}, \citenamefont {Galatali}, \citenamefont {Geyer}
  \emph {et~al.}}]{Urmson2008}%
  \BibitemOpen
  \bibfield  {author} {\bibinfo {author} {\bibfnamefont {C.}~\bibnamefont
  {Urmson}}, \bibinfo {author} {\bibfnamefont {J.}~\bibnamefont {Anhalt}},
  \bibinfo {author} {\bibfnamefont {D.}~\bibnamefont {Bagnell}}, \bibinfo
  {author} {\bibfnamefont {C.}~\bibnamefont {Baker}}, \bibinfo {author}
  {\bibfnamefont {R.}~\bibnamefont {Bittner}}, \bibinfo {author} {\bibfnamefont
  {M.}~\bibnamefont {Clark}}, \bibinfo {author} {\bibfnamefont
  {J.}~\bibnamefont {Dolan}}, \bibinfo {author} {\bibfnamefont
  {D.}~\bibnamefont {Duggins}}, \bibinfo {author} {\bibfnamefont
  {T.}~\bibnamefont {Galatali}}, \bibinfo {author} {\bibfnamefont
  {C.}~\bibnamefont {Geyer}}, \emph {et~al.},\ }\bibfield  {title} {\bibinfo
  {title} {Autonomous driving in urban environments: Boss and the urban
  challenge},\ }\href
  {https://onlinelibrary.wiley.com/doi/abs/10.1002/rob.20255} {\bibfield
  {journal} {\bibinfo  {journal} {Journal of Field Robotics}\ }\textbf
  {\bibinfo {volume} {25}},\ \bibinfo {pages} {425} (\bibinfo {year}
  {2008})}\BibitemShut {NoStop}%
\bibitem [{\citenamefont {Bostick}(1967)}]{Bostick1967}%
  \BibitemOpen
  \bibfield  {author} {\bibinfo {author} {\bibfnamefont {H.}~\bibnamefont
  {Bostick}},\ }\bibfield  {title} {\bibinfo {title} {A carbon dioxide laser
  radar system},\ }\href {https://doi.org/10.1109/JQE.1967.1074540} {\bibfield
  {journal} {\bibinfo  {journal} {IEEE Journal of Quantum Electronics}\
  }\textbf {\bibinfo {volume} {3}},\ \bibinfo {pages} {232} (\bibinfo {year}
  {1967})}\BibitemShut {NoStop}%
\bibitem [{\citenamefont {Pierrottet}\ \emph {et~al.}(2008)\citenamefont
  {Pierrottet}, \citenamefont {Amzajerdian}, \citenamefont {Petway},
  \citenamefont {Barnes}, \citenamefont {Lockard},\ and\ \citenamefont
  {Rubio}}]{Pierrottet2008}%
  \BibitemOpen
  \bibfield  {author} {\bibinfo {author} {\bibfnamefont {D.}~\bibnamefont
  {Pierrottet}}, \bibinfo {author} {\bibfnamefont {F.}~\bibnamefont
  {Amzajerdian}}, \bibinfo {author} {\bibfnamefont {L.}~\bibnamefont {Petway}},
  \bibinfo {author} {\bibfnamefont {B.}~\bibnamefont {Barnes}}, \bibinfo
  {author} {\bibfnamefont {G.}~\bibnamefont {Lockard}},\ and\ \bibinfo {author}
  {\bibfnamefont {M.}~\bibnamefont {Rubio}},\ }\bibfield  {title} {\bibinfo
  {title} {Linear {FMCW} laser radar for precision range and vector velocity
  measurements},\ }\href {https://doi.org/10.1557/PROC-1076-K04-06} {\bibfield
  {journal} {\bibinfo  {journal} {MRS Online Proceedings Library Archive}\
  }\textbf {\bibinfo {volume} {1076}} (\bibinfo {year} {2008})}\BibitemShut
  {NoStop}%
\bibitem [{\citenamefont {Martin}\ \emph {et~al.}(2018)\citenamefont {Martin},
  \citenamefont {Dodane}, \citenamefont {Leviandier}, \citenamefont {Dolfi},
  \citenamefont {Naughton}, \citenamefont {O’Brien}, \citenamefont
  {Spuessens}, \citenamefont {Baets}, \citenamefont {Lepage}, \citenamefont
  {Verheyen} \emph {et~al.}}]{Martin2018}%
  \BibitemOpen
  \bibfield  {author} {\bibinfo {author} {\bibfnamefont {A.}~\bibnamefont
  {Martin}}, \bibinfo {author} {\bibfnamefont {D.}~\bibnamefont {Dodane}},
  \bibinfo {author} {\bibfnamefont {L.}~\bibnamefont {Leviandier}}, \bibinfo
  {author} {\bibfnamefont {D.}~\bibnamefont {Dolfi}}, \bibinfo {author}
  {\bibfnamefont {A.}~\bibnamefont {Naughton}}, \bibinfo {author}
  {\bibfnamefont {P.}~\bibnamefont {O’Brien}}, \bibinfo {author}
  {\bibfnamefont {T.}~\bibnamefont {Spuessens}}, \bibinfo {author}
  {\bibfnamefont {R.}~\bibnamefont {Baets}}, \bibinfo {author} {\bibfnamefont
  {G.}~\bibnamefont {Lepage}}, \bibinfo {author} {\bibfnamefont
  {P.}~\bibnamefont {Verheyen}}, \emph {et~al.},\ }\bibfield  {title} {\bibinfo
  {title} {Photonic integrated circuit-based {FMCW} coherent {LiDAR}},\ }\href
  {https://doi.org/10.1109/JLT.2018.2840223} {\bibfield  {journal} {\bibinfo
  {journal} {Journal of Lightwave Technology}\ }\textbf {\bibinfo {volume}
  {36}},\ \bibinfo {pages} {4640} (\bibinfo {year} {2018})}\BibitemShut
  {NoStop}%
\bibitem [{\citenamefont {Poulton}\ \emph {et~al.}(2019)\citenamefont
  {Poulton}, \citenamefont {Byrd}, \citenamefont {Russo}, \citenamefont
  {Timurdogan}, \citenamefont {Khandaker}, \citenamefont {Vermeulen},\ and\
  \citenamefont {Watts}}]{Poulton2019}%
  \BibitemOpen
  \bibfield  {author} {\bibinfo {author} {\bibfnamefont {C.~V.}\ \bibnamefont
  {Poulton}}, \bibinfo {author} {\bibfnamefont {M.~J.}\ \bibnamefont {Byrd}},
  \bibinfo {author} {\bibfnamefont {P.}~\bibnamefont {Russo}}, \bibinfo
  {author} {\bibfnamefont {E.}~\bibnamefont {Timurdogan}}, \bibinfo {author}
  {\bibfnamefont {M.}~\bibnamefont {Khandaker}}, \bibinfo {author}
  {\bibfnamefont {D.}~\bibnamefont {Vermeulen}},\ and\ \bibinfo {author}
  {\bibfnamefont {M.~R.}\ \bibnamefont {Watts}},\ }\bibfield  {title} {\bibinfo
  {title} {Long-range lidar and free-space data communication with
  high-performance optical phased arrays},\ }\href
  {https://doi.org/10.1109/JSTQE.2019.2908555} {\bibfield  {journal} {\bibinfo
  {journal} {IEEE Journal of Selected Topics in Quantum Electronics}\ }\textbf
  {\bibinfo {volume} {25}},\ \bibinfo {pages} {1} (\bibinfo {year}
  {2019})}\BibitemShut {NoStop}%
\bibitem [{\citenamefont {Isaac}\ \emph {et~al.}(2019)\citenamefont {Isaac},
  \citenamefont {Song}, \citenamefont {Pinna}, \citenamefont {Coldren},\ and\
  \citenamefont {Klamkin}}]{Isaac2019}%
  \BibitemOpen
  \bibfield  {author} {\bibinfo {author} {\bibfnamefont {B.~J.}\ \bibnamefont
  {Isaac}}, \bibinfo {author} {\bibfnamefont {B.}~\bibnamefont {Song}},
  \bibinfo {author} {\bibfnamefont {S.}~\bibnamefont {Pinna}}, \bibinfo
  {author} {\bibfnamefont {L.~A.}\ \bibnamefont {Coldren}},\ and\ \bibinfo
  {author} {\bibfnamefont {J.}~\bibnamefont {Klamkin}},\ }\bibfield  {title}
  {\bibinfo {title} {Indium {Phosphide} {Photonic} {Integrated} {Circuit}
  {Transceiver} for {FMCW} {LiDAR}},\ }\href
  {https://doi.org/10.1109/JSTQE.2019.2911420} {\bibfield  {journal} {\bibinfo
  {journal} {IEEE Journal of Selected Topics in Quantum Electronics}\ }\textbf
  {\bibinfo {volume} {25}},\ \bibinfo {pages} {1} (\bibinfo {year} {2019})},\
  \bibinfo {note} {conference Name: IEEE Journal of Selected Topics in Quantum
  Electronics}\BibitemShut {NoStop}%
\bibitem [{\citenamefont {Sayyah}\ \emph {et~al.}(2022)\citenamefont {Sayyah},
  \citenamefont {Sarkissian}, \citenamefont {Patterson}, \citenamefont {Huang},
  \citenamefont {Efimov}, \citenamefont {Kim}, \citenamefont {Elliott},
  \citenamefont {Yang},\ and\ \citenamefont {Hammon}}]{Sayyah2022}%
  \BibitemOpen
  \bibfield  {author} {\bibinfo {author} {\bibfnamefont {K.}~\bibnamefont
  {Sayyah}}, \bibinfo {author} {\bibfnamefont {R.}~\bibnamefont {Sarkissian}},
  \bibinfo {author} {\bibfnamefont {P.}~\bibnamefont {Patterson}}, \bibinfo
  {author} {\bibfnamefont {B.}~\bibnamefont {Huang}}, \bibinfo {author}
  {\bibfnamefont {O.}~\bibnamefont {Efimov}}, \bibinfo {author} {\bibfnamefont
  {D.}~\bibnamefont {Kim}}, \bibinfo {author} {\bibfnamefont {K.}~\bibnamefont
  {Elliott}}, \bibinfo {author} {\bibfnamefont {L.}~\bibnamefont {Yang}},\ and\
  \bibinfo {author} {\bibfnamefont {D.}~\bibnamefont {Hammon}},\ }\bibfield
  {title} {\bibinfo {title} {Fully {Integrated} {FMCW} {LiDAR} {Optical}
  {Engine} on a {Single} {Silicon} {Chip}},\ }\href
  {https://doi.org/10.1109/JLT.2022.3145711} {\bibfield  {journal} {\bibinfo
  {journal} {Journal of Lightwave Technology}\ }\textbf {\bibinfo {volume}
  {40}},\ \bibinfo {pages} {2763} (\bibinfo {year} {2022})},\ \bibinfo {note}
  {conference Name: Journal of Lightwave Technology}\BibitemShut {NoStop}%
\bibitem [{\citenamefont {Liu}\ \emph {et~al.}(2022)\citenamefont {Liu},
  \citenamefont {Qiu}, \citenamefont {Ji}, \citenamefont {Lukashchuk},
  \citenamefont {He}, \citenamefont {Riemensberger}, \citenamefont {Hafermann},
  \citenamefont {Wang}, \citenamefont {Liu}, \citenamefont {Ronning},\ and\
  \citenamefont {Kippenberg}}]{Liu2022}%
  \BibitemOpen
  \bibfield  {author} {\bibinfo {author} {\bibfnamefont {Y.}~\bibnamefont
  {Liu}}, \bibinfo {author} {\bibfnamefont {Z.}~\bibnamefont {Qiu}}, \bibinfo
  {author} {\bibfnamefont {X.}~\bibnamefont {Ji}}, \bibinfo {author}
  {\bibfnamefont {A.}~\bibnamefont {Lukashchuk}}, \bibinfo {author}
  {\bibfnamefont {J.}~\bibnamefont {He}}, \bibinfo {author} {\bibfnamefont
  {J.}~\bibnamefont {Riemensberger}}, \bibinfo {author} {\bibfnamefont
  {M.}~\bibnamefont {Hafermann}}, \bibinfo {author} {\bibfnamefont {R.~N.}\
  \bibnamefont {Wang}}, \bibinfo {author} {\bibfnamefont {J.}~\bibnamefont
  {Liu}}, \bibinfo {author} {\bibfnamefont {C.}~\bibnamefont {Ronning}},\ and\
  \bibinfo {author} {\bibfnamefont {T.~J.}\ \bibnamefont {Kippenberg}},\
  }\bibfield  {title} {\bibinfo {title} {A photonic integrated circuit–based
  erbium-doped amplifier},\ }\href {https://doi.org/10.1126/science.abo2631}
  {\bibfield  {journal} {\bibinfo  {journal} {Science}\ }\textbf {\bibinfo
  {volume} {376}},\ \bibinfo {pages} {1309} (\bibinfo {year}
  {2022})}\BibitemShut {NoStop}%
\bibitem [{\citenamefont {Amann}(1992)}]{Amann1992}%
  \BibitemOpen
  \bibfield  {author} {\bibinfo {author} {\bibfnamefont {M.-C.}\ \bibnamefont
  {Amann}},\ }\bibfield  {title} {\bibinfo {title} {Phase noise limited
  resolution of coherent {LIDAR} using widely tunable laser diodes},\ }\href
  {https://doi.org/10.1049/el:19921077} {\bibfield  {journal} {\bibinfo
  {journal} {Electronics Letters}\ }\textbf {\bibinfo {volume} {28}},\ \bibinfo
  {pages} {1694} (\bibinfo {year} {1992})}\BibitemShut {NoStop}%
\bibitem [{\citenamefont {Karlsson}\ and\ \citenamefont
  {Olsson}(1999)}]{Karlsson1999}%
  \BibitemOpen
  \bibfield  {author} {\bibinfo {author} {\bibfnamefont {C.~J.}\ \bibnamefont
  {Karlsson}}\ and\ \bibinfo {author} {\bibfnamefont {F.~A.~A.}\ \bibnamefont
  {Olsson}},\ }\bibfield  {title} {\bibinfo {title} {Linearization of the
  frequency sweep of a frequency-modulated continuous-wave semiconductor laser
  radar and the resulting ranging performance},\ }\bibfield  {journal}
  {\bibinfo  {journal} {Applied Optics}\ }\textbf {\bibinfo {volume} {38}},\
  \href {https://doi.org/10.1364/AO.38.003376} {10.1364/AO.38.003376} (\bibinfo
  {year} {1999}),\ \bibinfo {note} {publisher: Optica Publishing
  Group}\BibitemShut {NoStop}%
\bibitem [{\citenamefont {Zhang}\ \emph {et~al.}(2019)\citenamefont {Zhang},
  \citenamefont {Buscaino}, \citenamefont {Wang}, \citenamefont {Shams-Ansari},
  \citenamefont {Reimer}, \citenamefont {Zhu}, \citenamefont {Kahn},\ and\
  \citenamefont {Lon{\v{c}}ar}}]{Zhang2019}%
  \BibitemOpen
  \bibfield  {author} {\bibinfo {author} {\bibfnamefont {M.}~\bibnamefont
  {Zhang}}, \bibinfo {author} {\bibfnamefont {B.}~\bibnamefont {Buscaino}},
  \bibinfo {author} {\bibfnamefont {C.}~\bibnamefont {Wang}}, \bibinfo {author}
  {\bibfnamefont {A.}~\bibnamefont {Shams-Ansari}}, \bibinfo {author}
  {\bibfnamefont {C.}~\bibnamefont {Reimer}}, \bibinfo {author} {\bibfnamefont
  {R.}~\bibnamefont {Zhu}}, \bibinfo {author} {\bibfnamefont {J.~M.}\
  \bibnamefont {Kahn}},\ and\ \bibinfo {author} {\bibfnamefont
  {M.}~\bibnamefont {Lon{\v{c}}ar}},\ }\bibfield  {title} {\bibinfo {title}
  {Broadband electro-optic frequency comb generation in a lithium niobate
  microring resonator},\ }\href {https://doi.org/10.1038/s41586-019-1008-7}
  {\bibfield  {journal} {\bibinfo  {journal} {Nature}\ }\textbf {\bibinfo
  {volume} {568}},\ \bibinfo {pages} {373} (\bibinfo {year}
  {2019})}\BibitemShut {NoStop}%
\bibitem [{\citenamefont {Behroozpour}\ \emph {et~al.}(2017)\citenamefont
  {Behroozpour}, \citenamefont {Sandborn}, \citenamefont {Quack}, \citenamefont
  {Seok}, \citenamefont {Matsui}, \citenamefont {Wu},\ and\ \citenamefont
  {Boser}}]{Behroozpour2016}%
  \BibitemOpen
  \bibfield  {author} {\bibinfo {author} {\bibfnamefont {B.}~\bibnamefont
  {Behroozpour}}, \bibinfo {author} {\bibfnamefont {P.~A.~M.}\ \bibnamefont
  {Sandborn}}, \bibinfo {author} {\bibfnamefont {N.}~\bibnamefont {Quack}},
  \bibinfo {author} {\bibfnamefont {T.-J.}\ \bibnamefont {Seok}}, \bibinfo
  {author} {\bibfnamefont {Y.}~\bibnamefont {Matsui}}, \bibinfo {author}
  {\bibfnamefont {M.~C.}\ \bibnamefont {Wu}},\ and\ \bibinfo {author}
  {\bibfnamefont {B.~E.}\ \bibnamefont {Boser}},\ }\bibfield  {title} {\bibinfo
  {title} {Electronic-{Photonic} {Integrated} {Circuit} for {3D}
  {Microimaging}},\ }\href {https://doi.org/10.1109/JSSC.2016.2621755}
  {\bibfield  {journal} {\bibinfo  {journal} {IEEE Journal of Solid-State
  Circuits}\ }\textbf {\bibinfo {volume} {52}},\ \bibinfo {pages} {161}
  (\bibinfo {year} {2017})},\ \bibinfo {note} {conference Name: IEEE Journal of
  Solid-State Circuits}\BibitemShut {NoStop}%
\bibitem [{\citenamefont {Tran}\ \emph {et~al.}(2019)\citenamefont {Tran},
  \citenamefont {Huang},\ and\ \citenamefont {Bowers}}]{Tran2019}%
  \BibitemOpen
  \bibfield  {author} {\bibinfo {author} {\bibfnamefont {M.~A.}\ \bibnamefont
  {Tran}}, \bibinfo {author} {\bibfnamefont {D.}~\bibnamefont {Huang}},\ and\
  \bibinfo {author} {\bibfnamefont {J.~E.}\ \bibnamefont {Bowers}},\ }\bibfield
   {title} {\bibinfo {title} {Tutorial on narrow linewidth tunable
  semiconductor lasers using {Si}/{III}-{V} heterogeneous integration},\ }\href
  {https://doi.org/10.1063/1.5124254} {\bibfield  {journal} {\bibinfo
  {journal} {APL Photonics}\ }\textbf {\bibinfo {volume} {4}},\ \bibinfo
  {pages} {111101} (\bibinfo {year} {2019})}\BibitemShut {NoStop}%
\bibitem [{\citenamefont {Jin}\ \emph {et~al.}(2021)\citenamefont {Jin},
  \citenamefont {Yang}, \citenamefont {Chang}, \citenamefont {Shen},
  \citenamefont {Wang}, \citenamefont {Leal}, \citenamefont {Wu}, \citenamefont
  {Gao}, \citenamefont {Feshali}, \citenamefont {Paniccia}, \citenamefont
  {Vahala},\ and\ \citenamefont {Bowers}}]{Jin2021}%
  \BibitemOpen
  \bibfield  {author} {\bibinfo {author} {\bibfnamefont {W.}~\bibnamefont
  {Jin}}, \bibinfo {author} {\bibfnamefont {Q.-F.}\ \bibnamefont {Yang}},
  \bibinfo {author} {\bibfnamefont {L.}~\bibnamefont {Chang}}, \bibinfo
  {author} {\bibfnamefont {B.}~\bibnamefont {Shen}}, \bibinfo {author}
  {\bibfnamefont {H.}~\bibnamefont {Wang}}, \bibinfo {author} {\bibfnamefont
  {M.~A.}\ \bibnamefont {Leal}}, \bibinfo {author} {\bibfnamefont
  {L.}~\bibnamefont {Wu}}, \bibinfo {author} {\bibfnamefont {M.}~\bibnamefont
  {Gao}}, \bibinfo {author} {\bibfnamefont {A.}~\bibnamefont {Feshali}},
  \bibinfo {author} {\bibfnamefont {M.}~\bibnamefont {Paniccia}}, \bibinfo
  {author} {\bibfnamefont {K.~J.}\ \bibnamefont {Vahala}},\ and\ \bibinfo
  {author} {\bibfnamefont {J.~E.}\ \bibnamefont {Bowers}},\ }\bibfield  {title}
  {\bibinfo {title} {Hertz-linewidth semiconductor lasers using {CMOS}-ready
  ultra-high- {Q} microresonators},\ }\href
  {https://doi.org/10.1038/s41566-021-00761-7} {\bibfield  {journal} {\bibinfo
  {journal} {Nature Photonics}\ ,\ \bibinfo {pages} {1}} (\bibinfo {year}
  {2021})}\BibitemShut {NoStop}%
\bibitem [{\citenamefont {Lihachev}\ \emph {et~al.}(2022)\citenamefont
  {Lihachev}, \citenamefont {Riemensberger}, \citenamefont {Weng},
  \citenamefont {Liu}, \citenamefont {Tian}, \citenamefont {Siddharth},
  \citenamefont {Snigirev}, \citenamefont {Shadymov}, \citenamefont {Voloshin},
  \citenamefont {Wang}, \citenamefont {He}, \citenamefont {Bhave},\ and\
  \citenamefont {Kippenberg}}]{Lihachev2022}%
  \BibitemOpen
  \bibfield  {author} {\bibinfo {author} {\bibfnamefont {G.}~\bibnamefont
  {Lihachev}}, \bibinfo {author} {\bibfnamefont {J.}~\bibnamefont
  {Riemensberger}}, \bibinfo {author} {\bibfnamefont {W.}~\bibnamefont {Weng}},
  \bibinfo {author} {\bibfnamefont {J.}~\bibnamefont {Liu}}, \bibinfo {author}
  {\bibfnamefont {H.}~\bibnamefont {Tian}}, \bibinfo {author} {\bibfnamefont
  {A.}~\bibnamefont {Siddharth}}, \bibinfo {author} {\bibfnamefont
  {V.}~\bibnamefont {Snigirev}}, \bibinfo {author} {\bibfnamefont
  {V.}~\bibnamefont {Shadymov}}, \bibinfo {author} {\bibfnamefont
  {A.}~\bibnamefont {Voloshin}}, \bibinfo {author} {\bibfnamefont {R.~N.}\
  \bibnamefont {Wang}}, \bibinfo {author} {\bibfnamefont {J.}~\bibnamefont
  {He}}, \bibinfo {author} {\bibfnamefont {S.~A.}\ \bibnamefont {Bhave}},\ and\
  \bibinfo {author} {\bibfnamefont {T.~J.}\ \bibnamefont {Kippenberg}},\
  }\bibfield  {title} {\bibinfo {title} {Low-noise frequency-agile photonic
  integrated lasers for coherent ranging},\ }\href
  {https://doi.org/10.1038/s41467-022-30911-6} {\bibfield  {journal} {\bibinfo
  {journal} {Nature Communications}\ }\textbf {\bibinfo {volume} {13}},\
  \bibinfo {pages} {3522} (\bibinfo {year} {2022})}\BibitemShut {NoStop}%
\bibitem [{\citenamefont {Li}\ \emph {et~al.}(2022{\natexlab{a}})\citenamefont
  {Li}, \citenamefont {Chang}, \citenamefont {Wu}, \citenamefont {Staffa},
  \citenamefont {Ling}, \citenamefont {Javid}, \citenamefont {Xue},
  \citenamefont {He}, \citenamefont {Lopez-rios}, \citenamefont {Morin},
  \citenamefont {Wang}, \citenamefont {Shen}, \citenamefont {Zeng},
  \citenamefont {Zhu}, \citenamefont {Vahala}, \citenamefont {Bowers},\ and\
  \citenamefont {Lin}}]{Li2022}%
  \BibitemOpen
  \bibfield  {author} {\bibinfo {author} {\bibfnamefont {M.}~\bibnamefont
  {Li}}, \bibinfo {author} {\bibfnamefont {L.}~\bibnamefont {Chang}}, \bibinfo
  {author} {\bibfnamefont {L.}~\bibnamefont {Wu}}, \bibinfo {author}
  {\bibfnamefont {J.}~\bibnamefont {Staffa}}, \bibinfo {author} {\bibfnamefont
  {J.}~\bibnamefont {Ling}}, \bibinfo {author} {\bibfnamefont {U.~A.}\
  \bibnamefont {Javid}}, \bibinfo {author} {\bibfnamefont {S.}~\bibnamefont
  {Xue}}, \bibinfo {author} {\bibfnamefont {Y.}~\bibnamefont {He}}, \bibinfo
  {author} {\bibfnamefont {R.}~\bibnamefont {Lopez-rios}}, \bibinfo {author}
  {\bibfnamefont {T.~J.}\ \bibnamefont {Morin}}, \bibinfo {author}
  {\bibfnamefont {H.}~\bibnamefont {Wang}}, \bibinfo {author} {\bibfnamefont
  {B.}~\bibnamefont {Shen}}, \bibinfo {author} {\bibfnamefont {S.}~\bibnamefont
  {Zeng}}, \bibinfo {author} {\bibfnamefont {L.}~\bibnamefont {Zhu}}, \bibinfo
  {author} {\bibfnamefont {K.~J.}\ \bibnamefont {Vahala}}, \bibinfo {author}
  {\bibfnamefont {J.~E.}\ \bibnamefont {Bowers}},\ and\ \bibinfo {author}
  {\bibfnamefont {Q.}~\bibnamefont {Lin}},\ }\bibfield  {title} {\bibinfo
  {title} {Integrated {Pockels} laser},\ }\href
  {https://doi.org/10.1038/s41467-022-33101-6} {\bibfield  {journal} {\bibinfo
  {journal} {Nature Communications}\ }\textbf {\bibinfo {volume} {13}},\
  \bibinfo {pages} {5344} (\bibinfo {year} {2022}{\natexlab{a}})}\BibitemShut
  {NoStop}%
\bibitem [{\citenamefont {Snigirev}\ \emph {et~al.}(2023)\citenamefont
  {Snigirev}, \citenamefont {Riedhauser}, \citenamefont {Lihachev},
  \citenamefont {Churaev}, \citenamefont {Riemensberger}, \citenamefont {Wang},
  \citenamefont {Siddharth}, \citenamefont {Huang}, \citenamefont {Möhl},
  \citenamefont {Popoff}, \citenamefont {Drechsler}, \citenamefont {Caimi},
  \citenamefont {Hönl}, \citenamefont {Liu}, \citenamefont {Seidler},\ and\
  \citenamefont {Kippenberg}}]{Snigirev2022}%
  \BibitemOpen
  \bibfield  {author} {\bibinfo {author} {\bibfnamefont {V.}~\bibnamefont
  {Snigirev}}, \bibinfo {author} {\bibfnamefont {A.}~\bibnamefont
  {Riedhauser}}, \bibinfo {author} {\bibfnamefont {G.}~\bibnamefont
  {Lihachev}}, \bibinfo {author} {\bibfnamefont {M.}~\bibnamefont {Churaev}},
  \bibinfo {author} {\bibfnamefont {J.}~\bibnamefont {Riemensberger}}, \bibinfo
  {author} {\bibfnamefont {R.~N.}\ \bibnamefont {Wang}}, \bibinfo {author}
  {\bibfnamefont {A.}~\bibnamefont {Siddharth}}, \bibinfo {author}
  {\bibfnamefont {G.}~\bibnamefont {Huang}}, \bibinfo {author} {\bibfnamefont
  {C.}~\bibnamefont {Möhl}}, \bibinfo {author} {\bibfnamefont
  {Y.}~\bibnamefont {Popoff}}, \bibinfo {author} {\bibfnamefont
  {U.}~\bibnamefont {Drechsler}}, \bibinfo {author} {\bibfnamefont
  {D.}~\bibnamefont {Caimi}}, \bibinfo {author} {\bibfnamefont
  {S.}~\bibnamefont {Hönl}}, \bibinfo {author} {\bibfnamefont
  {J.}~\bibnamefont {Liu}}, \bibinfo {author} {\bibfnamefont {P.}~\bibnamefont
  {Seidler}},\ and\ \bibinfo {author} {\bibfnamefont {T.~J.}\ \bibnamefont
  {Kippenberg}},\ }\bibfield  {title} {\bibinfo {title} {Ultrafast tunable
  lasers using lithium niobate integrated photonics},\ }\href
  {https://doi.org/10.1038/s41586-023-05724-2} {\bibfield  {journal} {\bibinfo
  {journal} {Nature}\ }\textbf {\bibinfo {volume} {615}},\ \bibinfo {pages}
  {411} (\bibinfo {year} {2023})}\BibitemShut {NoStop}%
\bibitem [{\citenamefont {Li}\ \emph {et~al.}(2022{\natexlab{b}})\citenamefont
  {Li}, \citenamefont {Wang}, \citenamefont {Lihachev}, \citenamefont {Tan},
  \citenamefont {Snigirev}, \citenamefont {Churaev}, \citenamefont {Kuznetsov},
  \citenamefont {Siddharth}, \citenamefont {Bereyhi}, \citenamefont
  {Riemensberger},\ and\ \citenamefont {Kippenberg}}]{Li2022a}%
  \BibitemOpen
  \bibfield  {author} {\bibinfo {author} {\bibfnamefont {Z.}~\bibnamefont
  {Li}}, \bibinfo {author} {\bibfnamefont {R.~N.}\ \bibnamefont {Wang}},
  \bibinfo {author} {\bibfnamefont {G.}~\bibnamefont {Lihachev}}, \bibinfo
  {author} {\bibfnamefont {Z.}~\bibnamefont {Tan}}, \bibinfo {author}
  {\bibfnamefont {V.}~\bibnamefont {Snigirev}}, \bibinfo {author}
  {\bibfnamefont {M.}~\bibnamefont {Churaev}}, \bibinfo {author} {\bibfnamefont
  {N.}~\bibnamefont {Kuznetsov}}, \bibinfo {author} {\bibfnamefont
  {A.}~\bibnamefont {Siddharth}}, \bibinfo {author} {\bibfnamefont {M.~J.}\
  \bibnamefont {Bereyhi}}, \bibinfo {author} {\bibfnamefont {J.}~\bibnamefont
  {Riemensberger}},\ and\ \bibinfo {author} {\bibfnamefont {T.~J.}\
  \bibnamefont {Kippenberg}},\ }\href
  {https://doi.org/10.48550/arXiv.2208.05556} {\bibinfo {title} {Tightly
  confining lithium niobate photonic integrated circuits and lasers}} (\bibinfo
  {year} {2022}{\natexlab{b}}),\ \bibinfo {note} {arXiv:2208.05556
  [physics]}\BibitemShut {NoStop}%
\bibitem [{\citenamefont {Rees}\ \emph {et~al.}(2020)\citenamefont {Rees},
  \citenamefont {Fan}, \citenamefont {Geskus}, \citenamefont {Klein},
  \citenamefont {Oldenbeuving}, \citenamefont {Slot},\ and\ \citenamefont
  {Boller}}]{Rees2020}%
  \BibitemOpen
  \bibfield  {author} {\bibinfo {author} {\bibfnamefont {A.~v.}\ \bibnamefont
  {Rees}}, \bibinfo {author} {\bibfnamefont {Y.}~\bibnamefont {Fan}}, \bibinfo
  {author} {\bibfnamefont {D.}~\bibnamefont {Geskus}}, \bibinfo {author}
  {\bibfnamefont {E.~J.}\ \bibnamefont {Klein}}, \bibinfo {author}
  {\bibfnamefont {R.~M.}\ \bibnamefont {Oldenbeuving}}, \bibinfo {author}
  {\bibfnamefont {P.~J. M. v.~d.}\ \bibnamefont {Slot}},\ and\ \bibinfo
  {author} {\bibfnamefont {K.-J.}\ \bibnamefont {Boller}},\ }\bibfield  {title}
  {\bibinfo {title} {Ring resonator enhanced mode-hop-free wavelength tuning of
  an integrated extended-cavity laser},\ }\href
  {https://doi.org/10.1364/OE.386356} {\bibfield  {journal} {\bibinfo
  {journal} {Optics Express}\ }\textbf {\bibinfo {volume} {28}},\ \bibinfo
  {pages} {5669} (\bibinfo {year} {2020})}\BibitemShut {NoStop}%
\bibitem [{\citenamefont {Lihachev}\ \emph {et~al.}(2023)\citenamefont
  {Lihachev}, \citenamefont {Bancora}, \citenamefont {Snigirev}, \citenamefont
  {Tian}, \citenamefont {Riemensberger}, \citenamefont {Shadymov},
  \citenamefont {Siddharth}, \citenamefont {Attanasio}, \citenamefont {Wang},
  \citenamefont {Visani}, \citenamefont {Voloshin}, \citenamefont {Bhave},\
  and\ \citenamefont {Kippenberg}}]{Bancora2022}%
  \BibitemOpen
  \bibfield  {author} {\bibinfo {author} {\bibfnamefont {G.}~\bibnamefont
  {Lihachev}}, \bibinfo {author} {\bibfnamefont {A.}~\bibnamefont {Bancora}},
  \bibinfo {author} {\bibfnamefont {V.}~\bibnamefont {Snigirev}}, \bibinfo
  {author} {\bibfnamefont {H.}~\bibnamefont {Tian}}, \bibinfo {author}
  {\bibfnamefont {J.}~\bibnamefont {Riemensberger}}, \bibinfo {author}
  {\bibfnamefont {V.}~\bibnamefont {Shadymov}}, \bibinfo {author}
  {\bibfnamefont {A.}~\bibnamefont {Siddharth}}, \bibinfo {author}
  {\bibfnamefont {A.}~\bibnamefont {Attanasio}}, \bibinfo {author}
  {\bibfnamefont {R.~N.}\ \bibnamefont {Wang}}, \bibinfo {author}
  {\bibfnamefont {D.}~\bibnamefont {Visani}}, \bibinfo {author} {\bibfnamefont
  {A.}~\bibnamefont {Voloshin}}, \bibinfo {author} {\bibfnamefont
  {S.}~\bibnamefont {Bhave}},\ and\ \bibinfo {author} {\bibfnamefont {T.~J.}\
  \bibnamefont {Kippenberg}},\ }\bibfield  {title} {\bibinfo {title} {Frequency
  agile photonic integrated external cavity laser},\ }\bibfield  {journal}
  {\bibinfo  {journal} {arXiv}\ }\href
  {https://doi.org/10.48550/arXiv.2303.00425} {10.48550/arXiv.2303.00425}
  (\bibinfo {year} {2023}),\ \bibinfo {note} {arXiv:2303.00425
  [physics]}\BibitemShut {NoStop}%
\bibitem [{\citenamefont {Feneyrou}\ \emph {et~al.}(2017)\citenamefont
  {Feneyrou}, \citenamefont {Leviandier}, \citenamefont {Minet}, \citenamefont
  {Pillet}, \citenamefont {Martin}, \citenamefont {Dolfi}, \citenamefont
  {Schlotterbeck}, \citenamefont {Rondeau}, \citenamefont {Lacondemine},
  \citenamefont {Rieu} \emph {et~al.}}]{Feneyrou2017}%
  \BibitemOpen
  \bibfield  {author} {\bibinfo {author} {\bibfnamefont {P.}~\bibnamefont
  {Feneyrou}}, \bibinfo {author} {\bibfnamefont {L.}~\bibnamefont
  {Leviandier}}, \bibinfo {author} {\bibfnamefont {J.}~\bibnamefont {Minet}},
  \bibinfo {author} {\bibfnamefont {G.}~\bibnamefont {Pillet}}, \bibinfo
  {author} {\bibfnamefont {A.}~\bibnamefont {Martin}}, \bibinfo {author}
  {\bibfnamefont {D.}~\bibnamefont {Dolfi}}, \bibinfo {author} {\bibfnamefont
  {J.-P.}\ \bibnamefont {Schlotterbeck}}, \bibinfo {author} {\bibfnamefont
  {P.}~\bibnamefont {Rondeau}}, \bibinfo {author} {\bibfnamefont
  {X.}~\bibnamefont {Lacondemine}}, \bibinfo {author} {\bibfnamefont
  {A.}~\bibnamefont {Rieu}}, \emph {et~al.},\ }\bibfield  {title} {\bibinfo
  {title} {Frequency-modulated multifunction lidar for anemometry, range
  finding, and velocimetry--1. theory and signal processing},\ }\href
  {https://doi.org/10.1364/AO.56.009663} {\bibfield  {journal} {\bibinfo
  {journal} {Applied optics}\ }\textbf {\bibinfo {volume} {56}},\ \bibinfo
  {pages} {9663} (\bibinfo {year} {2017})}\BibitemShut {NoStop}%
\bibitem [{\citenamefont {Axelsson}(2004)}]{Axelsson2004}%
  \BibitemOpen
  \bibfield  {author} {\bibinfo {author} {\bibfnamefont {S.}~\bibnamefont
  {Axelsson}},\ }\bibfield  {title} {\bibinfo {title} {Noise radar using random
  phase and frequency modulation},\ }\href
  {https://doi.org/10.1109/TGRS.2004.834589} {\bibfield  {journal} {\bibinfo
  {journal} {IEEE Transactions on Geoscience and Remote Sensing}\ }\textbf
  {\bibinfo {volume} {42}},\ \bibinfo {pages} {2370} (\bibinfo {year}
  {2004})}\BibitemShut {NoStop}%
\bibitem [{\citenamefont {Wang}(1982)}]{Wang1982}%
  \BibitemOpen
  \bibfield  {author} {\bibinfo {author} {\bibfnamefont {J.~Y.}\ \bibnamefont
  {Wang}},\ }\bibfield  {title} {\bibinfo {title} {Heterodyne laser radar-{SNR}
  from a diffuse target containing multiple glints},\ }\href
  {https://doi.org/10.1364/AO.21.000464} {\bibfield  {journal} {\bibinfo
  {journal} {Applied Optics}\ }\textbf {\bibinfo {volume} {21}},\ \bibinfo
  {pages} {464} (\bibinfo {year} {1982})}\BibitemShut {NoStop}%
\bibitem [{\citenamefont {Polman}\ \emph {et~al.}(1991)\citenamefont {Polman},
  \citenamefont {Jacobson}, \citenamefont {Eaglesham}, \citenamefont
  {Kistler},\ and\ \citenamefont {Poate}}]{polman_optical_1991}%
  \BibitemOpen
  \bibfield  {author} {\bibinfo {author} {\bibfnamefont {A.}~\bibnamefont
  {Polman}}, \bibinfo {author} {\bibfnamefont {D.~C.}\ \bibnamefont
  {Jacobson}}, \bibinfo {author} {\bibfnamefont {D.~J.}\ \bibnamefont
  {Eaglesham}}, \bibinfo {author} {\bibfnamefont {R.~C.}\ \bibnamefont
  {Kistler}},\ and\ \bibinfo {author} {\bibfnamefont {J.~M.}\ \bibnamefont
  {Poate}},\ }\bibfield  {title} {\bibinfo {title} {Optical doping of waveguide
  materials by {MeV} {Er} implantation},\ }\href
  {https://doi.org/10.1063/1.349234} {\bibfield  {journal} {\bibinfo  {journal}
  {Journal of Applied Physics}\ }\textbf {\bibinfo {volume} {70}},\ \bibinfo
  {pages} {3778} (\bibinfo {year} {1991})}\BibitemShut {NoStop}%
\bibitem [{\citenamefont {Pollnau}(2015)}]{pollnau_rare-earth-ion-doped_2015}%
  \BibitemOpen
  \bibfield  {author} {\bibinfo {author} {\bibfnamefont {M.}~\bibnamefont
  {Pollnau}},\ }\bibfield  {title} {\bibinfo {title} {Rare-earth-ion-doped
  channel waveguide lasers on silicon},\ }\href
  {https://doi.org/10.1109/JSTQE.2014.2351811} {\bibfield  {journal} {\bibinfo
  {journal} {IEEE Journal of Selected Topics in Quantum Electronics}\ }\textbf
  {\bibinfo {volume} {21}},\ \bibinfo {pages} {414} (\bibinfo {year} {2015})},\
  \bibinfo {note} {publisher: IEEE}\BibitemShut {NoStop}%
\bibitem [{\citenamefont {Lee}\ \emph {et~al.}(2021)\citenamefont {Lee},
  \citenamefont {Lee}, \citenamefont {Eovino}, \citenamefont {Park},
  \citenamefont {Liang}, \citenamefont {Lin}, \citenamefont {Yoo},\ and\
  \citenamefont {Yoo}}]{Lee2021}%
  \BibitemOpen
  \bibfield  {author} {\bibinfo {author} {\bibfnamefont {J.}~\bibnamefont
  {Lee}}, \bibinfo {author} {\bibfnamefont {K.-R.}\ \bibnamefont {Lee}},
  \bibinfo {author} {\bibfnamefont {B.~E.}\ \bibnamefont {Eovino}}, \bibinfo
  {author} {\bibfnamefont {J.~H.}\ \bibnamefont {Park}}, \bibinfo {author}
  {\bibfnamefont {L.~Y.}\ \bibnamefont {Liang}}, \bibinfo {author}
  {\bibfnamefont {L.}~\bibnamefont {Lin}}, \bibinfo {author} {\bibfnamefont
  {H.-J.}\ \bibnamefont {Yoo}},\ and\ \bibinfo {author} {\bibfnamefont
  {J.}~\bibnamefont {Yoo}},\ }\bibfield  {title} {\bibinfo {title} {A
  36-channel auto-calibrated front-end asic for a pmut-based miniaturized 3-d
  ultrasound system},\ }\href {https://doi.org/10.1109/JSSC.2021.3049560}
  {\bibfield  {journal} {\bibinfo  {journal} {IEEE Journal of Solid-State
  Circuits}\ }\textbf {\bibinfo {volume} {56}},\ \bibinfo {pages} {1910}
  (\bibinfo {year} {2021})}\BibitemShut {NoStop}%
\bibitem [{\citenamefont {Jung}\ \emph {et~al.}(2018)\citenamefont {Jung},
  \citenamefont {Rashid}, \citenamefont {Carpenter}, \citenamefont {Tekes},
  \citenamefont {Cowell}, \citenamefont {Freear}, \citenamefont {Degertekin},\
  and\ \citenamefont {Ghovanloo}}]{Jung2018}%
  \BibitemOpen
  \bibfield  {author} {\bibinfo {author} {\bibfnamefont {G.}~\bibnamefont
  {Jung}}, \bibinfo {author} {\bibfnamefont {M.~W.}\ \bibnamefont {Rashid}},
  \bibinfo {author} {\bibfnamefont {T.~M.}\ \bibnamefont {Carpenter}}, \bibinfo
  {author} {\bibfnamefont {C.}~\bibnamefont {Tekes}}, \bibinfo {author}
  {\bibfnamefont {D.~M.~J.}\ \bibnamefont {Cowell}}, \bibinfo {author}
  {\bibfnamefont {S.}~\bibnamefont {Freear}}, \bibinfo {author} {\bibfnamefont
  {F.~L.}\ \bibnamefont {Degertekin}},\ and\ \bibinfo {author} {\bibfnamefont
  {M.}~\bibnamefont {Ghovanloo}},\ }\bibfield  {title} {\bibinfo {title}
  {Single-chip reduced-wire active catheter system with programmable transmit
  beamforming and receive time-division multiplexing for intracardiac
  echocardiography},\ }in\ \href {https://doi.org/10.1109/ISSCC.2018.8310247}
  {\emph {\bibinfo {booktitle} {2018 IEEE International Solid - State Circuits
  Conference - (ISSCC)}}}\ (\bibinfo {year} {2018})\ pp.\ \bibinfo {pages}
  {188--190}\BibitemShut {NoStop}%
\bibitem [{\citenamefont {Sooksood}\ \emph {et~al.}(2011)\citenamefont
  {Sooksood}, \citenamefont {Noorsal}, \citenamefont {Becker},\ and\
  \citenamefont {Ortmanns}}]{Sooksood2011}%
  \BibitemOpen
  \bibfield  {author} {\bibinfo {author} {\bibfnamefont {K.}~\bibnamefont
  {Sooksood}}, \bibinfo {author} {\bibfnamefont {E.}~\bibnamefont {Noorsal}},
  \bibinfo {author} {\bibfnamefont {J.}~\bibnamefont {Becker}},\ and\ \bibinfo
  {author} {\bibfnamefont {M.}~\bibnamefont {Ortmanns}},\ }\bibfield  {title}
  {\bibinfo {title} {A neural stimulator front-end with arbitrary pulse shape,
  hv compliance and adaptive supply requiring 0.05mm2 in 0.35um hvcmos},\ }in\
  \href {https://doi.org/10.1109/ISSCC.2011.5746330} {\emph {\bibinfo
  {booktitle} {2011 IEEE International Solid-State Circuits Conference}}}\
  (\bibinfo {year} {2011})\ pp.\ \bibinfo {pages} {306--308}\BibitemShut
  {NoStop}%
\bibitem [{\citenamefont {Zhang}\ and\ \citenamefont
  {Zbinden}(2015)}]{Zhang2015}%
  \BibitemOpen
  \bibfield  {author} {\bibinfo {author} {\bibfnamefont {I.}~\bibnamefont
  {Zhang}}\ and\ \bibinfo {author} {\bibfnamefont {P.}~\bibnamefont
  {Zbinden}},\ }\bibfield  {title} {\bibinfo {title} {Advances in ingaas/inp
  single-photon detector systems for quantum communication},\ }\href
  {https://doi.org/https://doi.org/10.1038/lsa.2015.59} {\bibfield  {journal}
  {\bibinfo  {journal} {Light: Science \& Applications}\ }\textbf {\bibinfo
  {volume} {4}},\ \bibinfo {pages} {e286} (\bibinfo {year} {2015})}\BibitemShut
  {NoStop}%
\bibitem [{\citenamefont {Dragonas}\ \emph {et~al.}(2015)\citenamefont
  {Dragonas}, \citenamefont {Neretti}, \citenamefont {Sanjeevikumar},\ and\
  \citenamefont {Grandi}}]{Dragonas2015}%
  \BibitemOpen
  \bibfield  {author} {\bibinfo {author} {\bibfnamefont {F.~A.}\ \bibnamefont
  {Dragonas}}, \bibinfo {author} {\bibfnamefont {G.}~\bibnamefont {Neretti}},
  \bibinfo {author} {\bibfnamefont {P.}~\bibnamefont {Sanjeevikumar}},\ and\
  \bibinfo {author} {\bibfnamefont {G.}~\bibnamefont {Grandi}},\ }\bibfield
  {title} {\bibinfo {title} {High-voltage high-frequency arbitrary waveform
  multilevel generator for dbd plasma actuators},\ }\href
  {https://doi.org/10.1109/TIA.2015.2409262} {\bibfield  {journal} {\bibinfo
  {journal} {IEEE Transactions on Industry Applications}\ }\textbf {\bibinfo
  {volume} {51}},\ \bibinfo {pages} {3334} (\bibinfo {year}
  {2015})}\BibitemShut {NoStop}%
\bibitem [{\citenamefont {Palumbo}\ and\ \citenamefont
  {Pappalardo}(2010)}]{Palumbo2010}%
  \BibitemOpen
  \bibfield  {author} {\bibinfo {author} {\bibfnamefont {G.}~\bibnamefont
  {Palumbo}}\ and\ \bibinfo {author} {\bibfnamefont {D.}~\bibnamefont
  {Pappalardo}},\ }\bibfield  {title} {\bibinfo {title} {Charge pump circuits:
  An overview on design strategies and topologies},\ }\href
  {https://doi.org/10.1109/MCAS.2009.935695} {\bibfield  {journal} {\bibinfo
  {journal} {IEEE Circuits and Systems Magazine}\ }\textbf {\bibinfo {volume}
  {10}},\ \bibinfo {pages} {31} (\bibinfo {year} {2010})}\BibitemShut {NoStop}%
\bibitem [{\citenamefont {Ahn}\ and\ \citenamefont {Kim}(2007)}]{Ahn2007}%
  \BibitemOpen
  \bibfield  {author} {\bibinfo {author} {\bibfnamefont {T.-J.}\ \bibnamefont
  {Ahn}}\ and\ \bibinfo {author} {\bibfnamefont {D.~Y.}\ \bibnamefont {Kim}},\
  }\bibfield  {title} {\bibinfo {title} {Analysis of nonlinear frequency sweep
  in high-speed tunable laser sources using a self-homodyne measurement and
  hilbert transformation},\ }\href {https://doi.org/10.1364/AO.46.002394}
  {\bibfield  {journal} {\bibinfo  {journal} {Applied optics}\ }\textbf
  {\bibinfo {volume} {46}},\ \bibinfo {pages} {2394} (\bibinfo {year}
  {2007})}\BibitemShut {NoStop}%
\bibitem [{\citenamefont {Zhu}\ and\ \citenamefont {Zhu}(2019)}]{Zhu2019}%
  \BibitemOpen
  \bibfield  {author} {\bibinfo {author} {\bibfnamefont {Y.}~\bibnamefont
  {Zhu}}\ and\ \bibinfo {author} {\bibfnamefont {L.}~\bibnamefont {Zhu}},\
  }\bibfield  {title} {\bibinfo {title} {Narrow-linewidth, tunable external
  cavity dual-band diode lasers through
  {InP}/{GaAs}-{Si}$_{\textrm{3}}${N}$_{\textrm{4}}$ hybrid integration},\
  }\href {https://doi.org/10.1364/OE.27.002354} {\bibfield  {journal} {\bibinfo
   {journal} {Optics Express}\ }\textbf {\bibinfo {volume} {27}},\ \bibinfo
  {pages} {2354} (\bibinfo {year} {2019})}\BibitemShut {NoStop}%
\bibitem [{\citenamefont {Blaicher}\ \emph {et~al.}(2020)\citenamefont
  {Blaicher}, \citenamefont {Billah}, \citenamefont {Kemal}, \citenamefont
  {Hoose}, \citenamefont {Marin-Palomo}, \citenamefont {Hofmann}, \citenamefont
  {Kutuvantavida}, \citenamefont {Kieninger}, \citenamefont {Dietrich},
  \citenamefont {Lauermann}, \citenamefont {Wolf}, \citenamefont {Troppenz},
  \citenamefont {Moehrle}, \citenamefont {Merget}, \citenamefont {Skacel},
  \citenamefont {Witzens}, \citenamefont {Randel}, \citenamefont {Freude},\
  and\ \citenamefont {Koos}}]{Blaicher2020}%
  \BibitemOpen
  \bibfield  {author} {\bibinfo {author} {\bibfnamefont {M.}~\bibnamefont
  {Blaicher}}, \bibinfo {author} {\bibfnamefont {M.~R.}\ \bibnamefont
  {Billah}}, \bibinfo {author} {\bibfnamefont {J.}~\bibnamefont {Kemal}},
  \bibinfo {author} {\bibfnamefont {T.}~\bibnamefont {Hoose}}, \bibinfo
  {author} {\bibfnamefont {P.}~\bibnamefont {Marin-Palomo}}, \bibinfo {author}
  {\bibfnamefont {A.}~\bibnamefont {Hofmann}}, \bibinfo {author} {\bibfnamefont
  {Y.}~\bibnamefont {Kutuvantavida}}, \bibinfo {author} {\bibfnamefont
  {C.}~\bibnamefont {Kieninger}}, \bibinfo {author} {\bibfnamefont {P.-I.}\
  \bibnamefont {Dietrich}}, \bibinfo {author} {\bibfnamefont {M.}~\bibnamefont
  {Lauermann}}, \bibinfo {author} {\bibfnamefont {S.}~\bibnamefont {Wolf}},
  \bibinfo {author} {\bibfnamefont {U.}~\bibnamefont {Troppenz}}, \bibinfo
  {author} {\bibfnamefont {M.}~\bibnamefont {Moehrle}}, \bibinfo {author}
  {\bibfnamefont {F.}~\bibnamefont {Merget}}, \bibinfo {author} {\bibfnamefont
  {S.}~\bibnamefont {Skacel}}, \bibinfo {author} {\bibfnamefont
  {J.}~\bibnamefont {Witzens}}, \bibinfo {author} {\bibfnamefont
  {S.}~\bibnamefont {Randel}}, \bibinfo {author} {\bibfnamefont
  {W.}~\bibnamefont {Freude}},\ and\ \bibinfo {author} {\bibfnamefont
  {C.}~\bibnamefont {Koos}},\ }\bibfield  {title} {\bibinfo {title} {Hybrid
  multi-chip assembly of optical communication engines by in situ {3D}
  nano-lithography},\ }\href {https://doi.org/10.1038/s41377-020-0272-5}
  {\bibfield  {journal} {\bibinfo  {journal} {Light: Science \& Applications}\
  }\textbf {\bibinfo {volume} {9}},\ \bibinfo {pages} {1} (\bibinfo {year}
  {2020})}\BibitemShut {NoStop}%
\bibitem [{\citenamefont {Pfeiffer}\ \emph
  {et~al.}(2018{\natexlab{a}})\citenamefont {Pfeiffer}, \citenamefont
  {Herkommer}, \citenamefont {Liu}, \citenamefont {Morais}, \citenamefont
  {Zervas}, \citenamefont {Geiselmann},\ and\ \citenamefont
  {Kippenberg}}]{Pfeiffer2018d}%
  \BibitemOpen
  \bibfield  {author} {\bibinfo {author} {\bibfnamefont {M.~H.~P.}\
  \bibnamefont {Pfeiffer}}, \bibinfo {author} {\bibfnamefont {C.}~\bibnamefont
  {Herkommer}}, \bibinfo {author} {\bibfnamefont {J.}~\bibnamefont {Liu}},
  \bibinfo {author} {\bibfnamefont {T.}~\bibnamefont {Morais}}, \bibinfo
  {author} {\bibfnamefont {M.}~\bibnamefont {Zervas}}, \bibinfo {author}
  {\bibfnamefont {M.}~\bibnamefont {Geiselmann}},\ and\ \bibinfo {author}
  {\bibfnamefont {T.~J.}\ \bibnamefont {Kippenberg}},\ }\bibfield  {title}
  {\bibinfo {title} {Photonic damascene process for low-loss, high-confinement
  silicon nitride waveguides},\ }\href
  {https://doi.org/10.1109/JSTQE.2018.2808258} {\bibfield  {journal} {\bibinfo
  {journal} {IEEE Journal of Selected Topics in Quantum Electronics}\ }\textbf
  {\bibinfo {volume} {24}},\ \bibinfo {pages} {1} (\bibinfo {year}
  {2018}{\natexlab{a}})}\BibitemShut {NoStop}%
\bibitem [{\citenamefont {Pfeiffer}\ \emph
  {et~al.}(2018{\natexlab{b}})\citenamefont {Pfeiffer}, \citenamefont {Liu},
  \citenamefont {Raja}, \citenamefont {Morais}, \citenamefont {Ghadiani},\ and\
  \citenamefont {Kippenberg}}]{Pfeiffer2018}%
  \BibitemOpen
  \bibfield  {author} {\bibinfo {author} {\bibfnamefont {M.~H.~P.}\
  \bibnamefont {Pfeiffer}}, \bibinfo {author} {\bibfnamefont {J.}~\bibnamefont
  {Liu}}, \bibinfo {author} {\bibfnamefont {A.~S.}\ \bibnamefont {Raja}},
  \bibinfo {author} {\bibfnamefont {T.}~\bibnamefont {Morais}}, \bibinfo
  {author} {\bibfnamefont {B.}~\bibnamefont {Ghadiani}},\ and\ \bibinfo
  {author} {\bibfnamefont {T.~J.}\ \bibnamefont {Kippenberg}},\ }\bibfield
  {title} {\bibinfo {title} {Ultra-smooth silicon nitride waveguides based on
  the damascene reflow process: fabrication and loss origins},\ }\href
  {https://doi.org/10.1364/OPTICA.5.000884} {\bibfield  {journal} {\bibinfo
  {journal} {Optica}\ }\textbf {\bibinfo {volume} {5}},\ \bibinfo {pages} {884}
  (\bibinfo {year} {2018}{\natexlab{b}})}\BibitemShut {NoStop}%
\end{thebibliography}

%


\newpage
\section{Methods}

\subsection{Vernier ring laser operation}

In our experiments, the laser operated at a temperature of 26~C$^o$ (thermistor resistance 9700 OHm) and RSOA current of 340~mA, resulting in 3~mW output optical power.
We applied around 30~mW of electrical power to microheaters to align resonances of \SiN microrings.
We do not have on-chip phase shifter section in our laser design, and the tuning range was empirically limited to 3~GHz while the RSOA current was fixed.

\subsection{EDWA sample fabrication}
\noindent
Integrated \SiN spiral waveguides were fabricated with the photonic damascene process~\cite{Pfeiffer2018d}. 
The preform for waveguide structures and filler patterns for stress release was then defined by deep ultra-violet photolithography and transferred into the thermal oxide layer using reactive ion etching (RIE).
Thermal treatment at \SI{1250}{\degreeCelsius} was then applied to reflow the silicon oxide, reducing the roughness caused by the RIE etching \cite{Pfeiffer2018}, before the preform recesses were filled with stoichiometric \SiN by LPCVD.
An etchback process was performed to roughly planarize the wafer surface and remove the excess \SiN material.
Chemical mechanical polishing (CMP) was then applied to reach the desired waveguide thickness, and create a top surface with sub-nanometer root-mean-square roughness.
Using this process,  700~nm-thick \SiN waveguides buried in a wet oxide cladding but without top cladding were created, which allows for direct Erbium implantation into the waveguide core.
The Erbium ion beam energy of 0.955~MeV, 1.416~MeV and 2~MeV and the corresponding fluence of $2.34 \times 10^{15}$, $3.17\times 10^{15}$ and $4.5\times10^{15}$~ions/cm\textsuperscript{2}, respectively, were consecutively applied to the separated passive \SiN photonic chips.
This process can implant Erbium ions into the \SiN waveguides ($0.7\times2.1~\mathrm{\mu m^2}$ cross-section) with a maximum doping depth of ca. 400~nm from the top surface and achieve an overlap factor of ca. 50~\%.
The doped \SiN was then annealed at \SI{1000}{\degreeCelsius} in O\textsubscript{2} under atmosphere pressure for 1 hour to heal the implantation defects and optically activate the doped Erbium ions.
The measured lifetime of the first excited state of doped Erbium ions is ca.  3.4 ms.
Higher annealing temperature could lead to Erbium ion precipitations in the silica cladding \cite{polman_optical_1991}.
The Erbium ion implantation of the \SiN photonic chips was performed at the University of Surrey Ion Beam Centre via commercial service.

\subsection{HV-AWG ASIC operation}
The design was fabricated in a standard 130-nm SiGe BiCMOS technology, where only CMOS transistors were used. The chip measures 1.17-1.07 mm$^2$. The total active area of the design, i. e. excluding decoupling caps and IO pads, is approximately 35,000 $\mu$m$^2$. The charge pump is implemented with diode connected isolated thick oxide NMOS transistors and metal-insulator-metal (MiM) capacitors. 
$V_\mathrm{in}$ and $V_\mathrm{cascode}$ were biased at $V_\mathrm{DD}$= 3.3V. The 15-stage charge pump has three diode-connected transistors in each stage, so as to set $V_\mathrm{discharge}$ to $V_\mathrm{out}$ – 45$V_\mathrm{th,n}$ as $V_\mathrm{out}$ rises during the charging phase. A cascode transistor is also introduced at $V_\mathrm{discharge}$, so that a voltage up to 2$V_\mathrm{DD}$ can be tolerated at this node. This ensures that $V_\mathrm{out}$ can safely reach more than 20 volts without reaching breakdown of the discharge transistors, given that the capacitors and NMOS-to-bulk isolation are within breakdown limits as well. For the measurements presented, the output of the CMOS chip was fed to an external unity-gain voltage buffer whose output was measured and/or used to drive the actuators. Load capacitance at chip output was measured at 26 pF. The printed circuit board used for controlling the ASIC is shown in SI Fig.2. 

\bigskip
\section*{SUPPLEMENTARY INFORMATION FIGURES} 
\renewcommand{\figurename}{\textbf{SI Fig.}}
\setcounter{figure}{0}
\begin{figure*}[!htbp]  
\centering
	\includegraphics[width=0.75\linewidth]{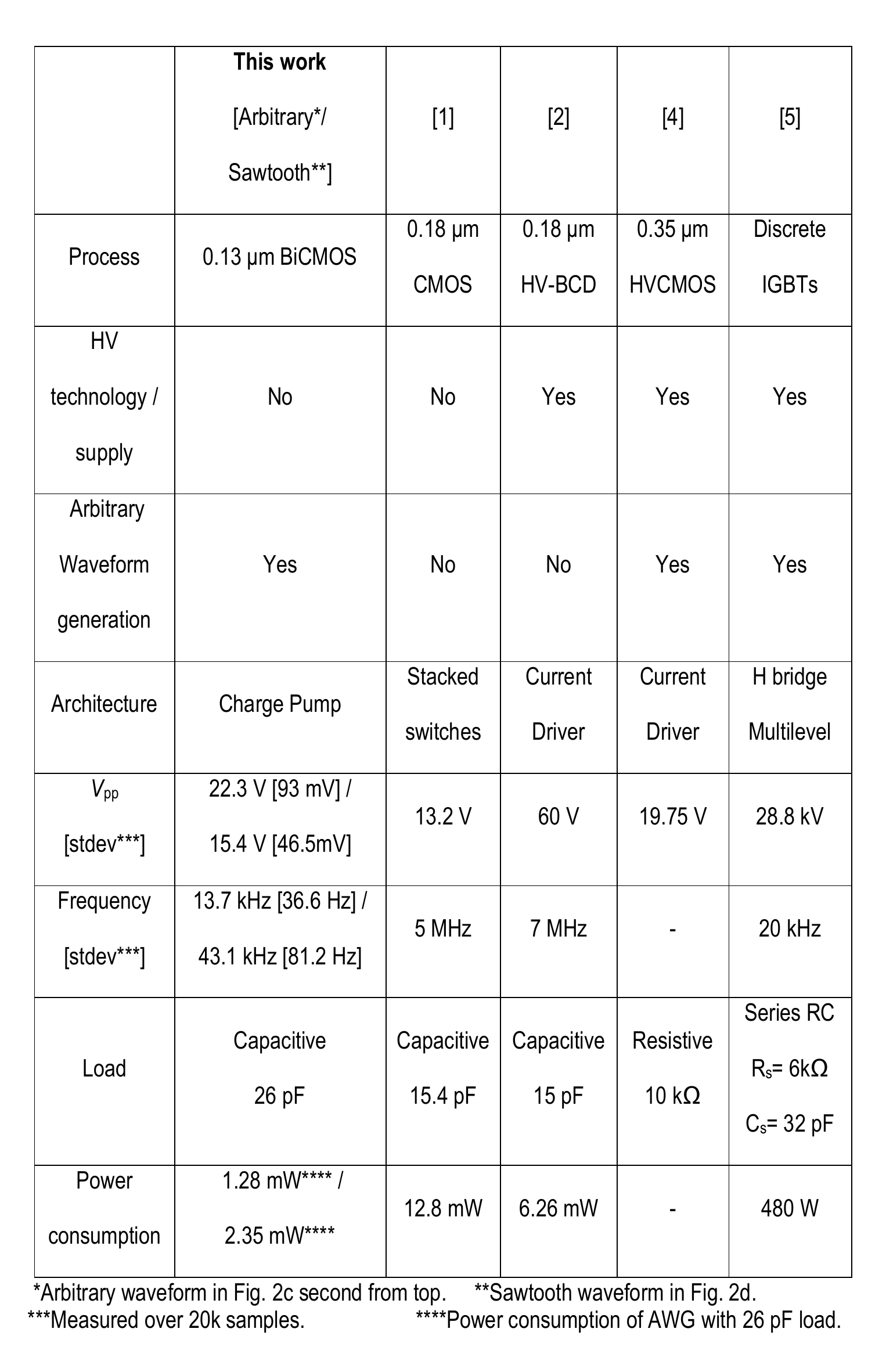}
	\caption{\textbf{HV-AWG comparison table}
	}
	\label{fig_SI_table}
\end{figure*}

\newpage

\begin{figure*}[!htbp]  
\centering
	\includegraphics[width=0.9\linewidth]{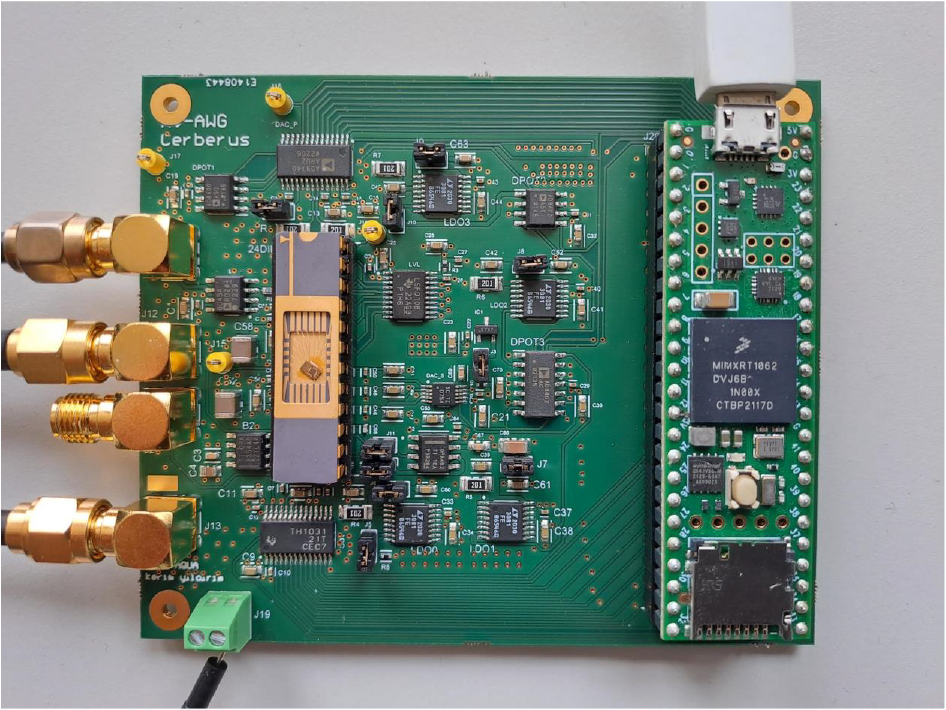}
	\caption{\textbf{Printed circuit board (PCB) of the HV-AWG}
 The output voltage is connected to a unity-gain voltage buffer on the PCB through the package, and the buffer output is used to drive the SMA connector that is used for the measurements.
	}
	\label{fig_SI_pcb}
\end{figure*}

\newpage
\begin{figure*}[!htbp] 
\centering
	\includegraphics[width=0.9\linewidth]{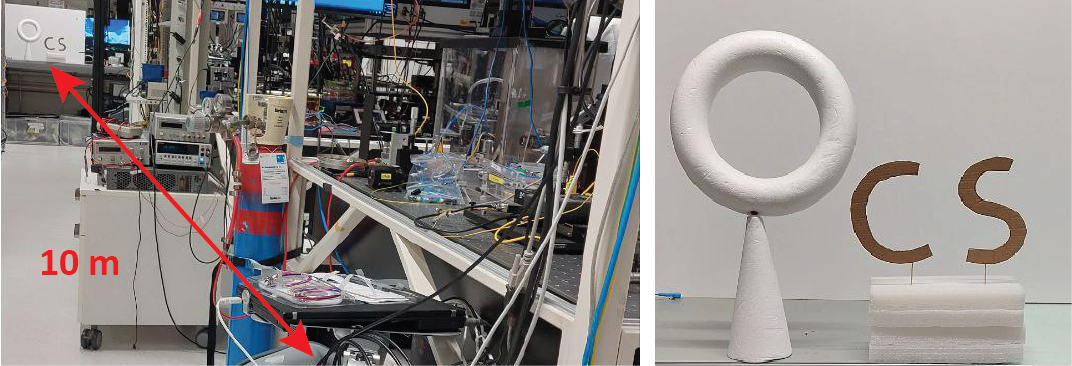}
	\caption{\textbf{Imaging scene}
	}
	\label{fig_SI_scene}
\end{figure*}

\end{document}